\documentclass[10pt, doublecolumn]{IEEEtran}
\usepackage{epsfig,latexsym}
\usepackage{float}
\usepackage{indentfirst}
\usepackage{amsmath}
\usepackage{bm}
\usepackage{amssymb}
\usepackage{times}
\usepackage{enumitem}
\usepackage{algorithm}
\usepackage{algorithmic}
\usepackage{algorithm}
\usepackage{cite,hyperref}
\usepackage{lastpage}
\usepackage{color} 
\usepackage{amsthm}
\usepackage{array}

\def\BibTeX{{\rm B\kern-.05em{\sc i\kern-.025em b}\kern-.08em
    T\kern-.1667em\lower.7ex\hbox{E}\kern-.125emX}}
    \expandafter\def\expandafter\normalsize\expandafter{%
    \normalsize%
    \setlength\abovedisplayskip{4pt}%
    \setlength\belowdisplayskip{4pt}%
    \setlength\abovedisplayshortskip{2pt}%
    \setlength\belowdisplayshortskip{2pt}%
}
\begin{document}
\title{Multigroup Multicast Design for Pinching-Antenna Systems: Waveguide-Division or Waveguide-Multiplexing?}
\author{Shan Shan, Chongjun Ouyang, Yong Li, and Yuanwei Liu\vspace{-10pt}
\thanks{Shan Shan and Yong Li are with the School of Information and Communication Engineering, Beijing University of Posts and Telecommunications, Beijing 100876, China (e-mail: \{shan.shan, liyong\}@bupt.edu.cn). Chongjun Ouyang is with the School of Electronic Engineering and Computer Science, Queen Mary University of London, London E1 4NS, U.K. (e-mail: c.ouyang@qmul.ac.uk). Yuanwei Liu is with the Department of Electrical and Electronic Engineering, The University of Hong Kong, Hong Kong (e-mail: yuanwei@hku.hk).}
}
\maketitle
\begin{abstract}
This article addresses the design of multigroup multicast communications in the pinching-antenna system (PASS). A PASS-enabled multigroup transmission framework is proposed to maximize multicast rates under a couple of transmission architectures: waveguide-division (WD) and waveguide-multiplexing (WM). 1) For WD, an element-wise sequential optimization strategy is proposed for \emph{pinching beamforming}, i.e., optimizing the activated positions of pinching antennas along dielectric waveguides. Meanwhile, a log-sum-exp projected gradient descent algorithm is proposed for transmit power allocation across waveguides. 2) For WM, a majorization-minimization (MM)-based framework is proposed to tackle the problem's non-smoothness and non-convexity. On this basis, a low-complexity element-wise sequential optimization method is developed for pinching beamforming using the MM surrogate objective. Furthermore, the optimal transmit beamformer structure is derived from the MM surrogate objective using the Lagrange duality, with an efficient transmit beamforming algorithm proposed using projected adaptive gradient descent. Numerical results demonstrate that: i) both WD and WM architectures in PASS achieve significant multicast rate improvements over conventional MIMO techniques, especially for systems with large service areas; ii) WM is more robust than WD in dense deployments, while WD excels when user groups are spatially separated.
\end{abstract}

\begin{IEEEkeywords}
Beamforming, minimum maximization, multigroup multicast communication, pinching-antenna systems.
\end{IEEEkeywords}

\section{Introduction}
The evolution of wireless communication systems has been driven by continuous efforts to enhance spectral efficiency, coverage, and reliability. Traditional multiple-input multiple-output (MIMO) systems have significantly contributed to these advancements by exploiting spatial diversity and multiplexing gains~\cite{6542746}. However, the fixed configuration of conventional antenna deployments leads to limitations on the adaptability of wireless channel, particularly in dynamic environments with user mobility and time-varying propagation conditions.
To address these challenges, the concept of flexible-antenna systems has emerged, which encompasses technologies such as reconfigurable intelligent surfaces (RISs)~\cite{9424177}, fluid-antennas~\cite{10753482, 9264694}, and movable-antennas~\cite{10286328, 10906511}. These systems introduce the capability to reconfigure the wireless propagation environment dynamically, which improves wireless link quality and combats small-scale fading. Specifically, RISs manipulate the phase and amplitude of incident signals through passive reflective elements to reconfigure the propagation environment and enhance the received signal power~\cite{9424177}. 
Fluid-antenna and movable-antenna systems offer the ability to dynamically adjust antenna positions and their electromagnetic properties, which facilitates the formation of favorable spatial channels for MIMO communications~\cite{10753482, 9264694, 10286328, 10906511}.

Despite their innovative approaches, existing flexible-antenna technologies are subject to inherent constraints. In particular, RISs often suffer from the double-fading effect due to their reliance on reflected paths, which leads to increased path loss compared to direct line-of-sight (LoS) links. Fluid-antennas and movable-antennas, which offer antenna positional adaptability, are typically constrained to movements within a few wavelengths. This property limits their flexibility and effectiveness in mitigating large-scale path loss and LoS blockages. Moreover, these systems typically adopt a fixed antenna configuration--such as the number of antenna elements--that cannot be reconfigured once deployed, which limits their scalability and adaptability under different network scenarios.

\subsection{Pinching Antennas}
Recently, the \emph{Pinching-Antenna SyStem} (PASS) has been proposed and experimentally demonstrated by NTT DOCOMO as a novel flexible-antenna technology that addresses the aforementioned limitations~\cite{fukuda2022pinching}. PASS employs a dielectric waveguide as the transmission medium with low in-waveguide propagation loss, and its aperture length spans from a few meters to tens of meters. Along the waveguide, small dielectric elements, termed \emph{pinching antennas} (PAs), can be dynamically attached or detached. Radio waves are transmitted or received from the activated (pinched) locations along the waveguide.
A key advantage of PASS lies in its ability to establish \emph{``near-wired''} communications, as the waveguide can be arbitrarily long. This property enables the establishment of strong LoS links to individual users, which can effectively mitigate large-scale path loss and avoid LoS blockage~\cite{liu2025pass, 10945421}. Each PA can be independently activated or deactivated at any point along the waveguide. This capability enables dynamic reconfiguration of the antenna array geometry as well as the wireless channel. This mechanism facilitates a highly flexible and scalable deployment strategy, which we refer to as \emph{pinching beamforming} \cite{liu2025pass, 10945421}.

Conceptually, PASS can be regarded as a practical realization of fluid- or movable-antenna systems, but with enhanced flexibility and scalability compared to traditional architectures. In recognition of NTT DOCOMO's contribution, we adopt the term \emph{``PASS''} throughout this article. PASS is also well aligned with the emerging vision of surface-wave communication superhighways \cite{9210135}, which aim to exploit in-waveguide propagation through reconfigurable waveguides to minimize path loss and enhance signal delivery efficiency \cite{10643519, 10742352}.

\subsection{Prior Works}
Driven by these promising features, PASS has recently attracted considerable research attention. As a fundamental step toward understanding PASS, modelling the power radiation and propagation behavior of PASS has been considered in several recent studies. 
In particular, the authors in~\cite{wang2025modeling} established the fundamental power radiation model for PASS using electromagnetic coupling theory. The authors in~\cite{xu2025pass} proposed a novel adjustable power radiation model, where power radiation ratios of the PAs can be flexibly controlled by tuning the spacing between PAs and waveguides. Moreover, the impact of in-waveguide power propagation loss in the dielectric waveguide was discussed in~\cite{10912473}.

In parallel with those modeling efforts, many studies have analyzed the performance of PASS. Specifically, the ergodic rate gain of PASS over conventional fixed-location antenna systems was analyzed in~\cite{10945421}. The closed-form expressions for outage probability and average rate for single-PA based PASS were derived in~\cite{10976621}. In addition, the array gain of multiple-PA based PASS was characterized in~\cite{ouyang2025array}.

Building upon the aforementioned theoretical foundations, PASS have been applied to various wireless communication scenarios. For the single-waveguide employment, uplink transmissions was considered in~\cite{10909665, hou2025uplink} to enhance fairness by maximizing the minimum user rate through pinching beamforming design. For downlink scenarios, the authors in~\cite{10896748} addressed the optimization of PA positioning to maximize the channel gain. A closed-form expression for the placement of PAs was derived in~\cite{11016750} to maximize the sum rate of users.

To further explore the potential of PASS in spatial multiplexing and multiple access, the multiple-waveguide architecture has been considered for multiuser scenarios~\cite{bereyhi2025downlink, guo2025gpass, xu2025joint, wang2025modeling}. To support spatial multiplexing, a joint baseband and pinching beamforming optimization framework was developed in~\cite{wang2025modeling} to minimize transmit power. A hybrid beamforming framework was proposed in~\cite{bereyhi2025downlink} to optimize joint beamforming via fractional programming. The authors in~\cite{guo2025gpass} developed a staged graph-based learning architecture to sequentially learn pinching and transmit beamformer. In~\cite{xu2025joint}, both optimization-based and learning-based methods were proposed for sum-rate maximization. To enable multiuser access via dedicated waveguides, a waveguide division multiple access technique was introduced in~\cite{zhao2025wdma}, which underpins multiuser communication by leveraging the near-orthogonality of channels in free-space propagation.

\subsection{Motivations and Contributions}
The aforementioned studies have validated the effectiveness of PASS in enhancing the performance of wireless communications. However, existing contributions have mainly focused on exploring the performance benefits of deploying PASS in unicast transmission. In contrast, the application of PASS in multigroup \emph{multicast} systems--where different groups of users receive individual multicast messages--remains largely unexplored. Compared to unicast, multicast communications imposes stricter fairness requirements, as each group's rate is limited by its worst-case user. The high flexible reconfigurability of PASS can effectively reduce the average distance between each user and its nearest antenna element to establish strong LoS links for all users simultaneously, which makes PASS particularly well-suited for multicast communications.

Recall that each waveguide can carry only a single data stream, employing multiple-waveguide is essential in multigroup multicast systems to enable the simultaneous delivery of distinct group messages. A couple of representative multiple-waveguide based transmission architectures have been proposed: 1) waveguide-division (WD), in which each waveguide transmits an independent data stream; and 2) waveguide-multiplexing (WM), where all waveguides jointly transmit a linear combination of multiple streams via baseband transmit beamforming~\cite{liu2025pass}.
While both architectures have shown potential in unicast scenarios, their applicability and performance trade-offs in multicast systems have not been systematically studied. This article aims to fill this gap by addressing the following key question: \emph{Given the architectural options of WD and WM, which transmission design is more effective for multigroup multicast communications across diverse deployment scenarios?} The main contributions are summarized as follows.

\begin{itemize}
	\item We refine the power radiation model of PASS by incorporating both in-waveguide propagation loss and power radiation coefficient model. Building upon this characterization, we formulate a PASS-enabled multigroup multicast transmission framework that aims to maximize the overall multicast rate--defined as the sum of the minimum achievable rates within each user group--to ensure rate fairness and robust multigroup performance. Furthermore, we consider both waveguide-division (WD) and waveguide-multiplexing (WM) architectures to facilitate the multicast transmission.
	\item For the WD architecture, we formulate a joint power allocation and pinching beamforming problem to maximize the multicast rate. To tackle the non-convexity of the pinching beamforming, we propose an element-wise sequential optimization strategy that iteratively updates each PA location. For power allocation across waveguides, we propose a log-sum-exp projected gradient descent (LSE-PGD) algorithm, which ensures stable convergence and tractable complexity.
	\item For the WM architecture, we propose an alternating optimization (AO) algorithm based on the majorization-minimization (MM) framework to address the non-smooth and non-convex joint transmit and pinching beamforming design. In particular, we develop a low-complexity MM-based element-wise sequential method to optimize the pinching beamforming. Furthermore, leveraging the MM surrogate function, we analytically derive the optimal structure of the transmit beamformer and subsequently propose a projected adaptive gradient descent (MM-PAGD) algorithm to optimize the transmit beamformer.
	\item We present numerical results to validate the performance advantage of PASS and the effectiveness of the proposed algorithms. The results demonstrate that: i) PASS achieves significantly higher multicast rates compared to conventional fixed-location antenna systems, including conventional MIMO and massive MIMO systems, especially as the spatial coverage increase; ii) WM is better suited to dense user deployments, whereas WD is more appropriate for scenarios with spatially distributed user groups; and iii) increasing the number of waveguides and PAs further improves the multicast rate of PASS. These results confirm the scalability and potential of PASS for physical-layer multigroup multicast. 
\end{itemize}

\subsection{Organization and Notations}
The remainder of this paper is organized as follows. Section~\ref{System_Model} introduces the system model and formulates the sum minimum-rate maximization problem. 
Sections~\ref{WD_architecture} and \ref{WM_architecture} present the multigroup multicast design for the WD and WM architectures, respectively. Numerical results are presented in Section~\ref{simulation}. Finally, Section~\ref{conclusion} concludes this paper.
\subsubsection*{Notations}
Scalars, vectors, and matrices are represented by regular, bold lowercase, and  bold uppercase letters, respectively. 
The sets of complex numbers, real numbers, and non-negative real numbers are denoted by $\mathbb{C}$, $\mathbb{R}$, and $\mathbb{R}_{+}$, respectively. The inverse, conjugate, transpose, conjugate transpose, and trace operators are denoted by $(\cdot)^{-1}$, $(\cdot)^*$, $(\cdot)^{\rm T}$, $(\cdot)^{\rm H}$, and $\mathrm{Tr}(\cdot)$, respectively. For a vector $\mathbf{x}$, $[\mathbf{x}]_i$ denotes its $i$th element. ${\rm{Diag}}(\mathbf{x})$ denotes the diagonal matrix with the diagonal elements formed by the vector $\mathbf{x}$. ${\rm{BlkDiag}}([{\mathbf{A}};{\mathbf{B}}])$ denotes the block-diagonal matrix formed by the matrices $\mathbf{A}$ and $\mathbf{B}$. ${\mathcal C}{\mathcal N}(a, b^2)$ denotes a circularly symmetric complex Gaussian distribution with mean $a$ and variance $b^2$. The statistical expectation operator is represented by ${\mathbb{E}}\{\cdot\}$. The absolute value and Euclidean norm are denoted by $|\cdot|$ and $\|\cdot\|$, respectively. The real part of a complex number is denoted by $\Re \{\cdot\}$. The big-O notation is denoted by ${\mathcal O}(\cdot)$. 

\section{System Model}\label{System_Model}
\begin{figure}[!t]
\centering
\includegraphics[height=0.28\textwidth]{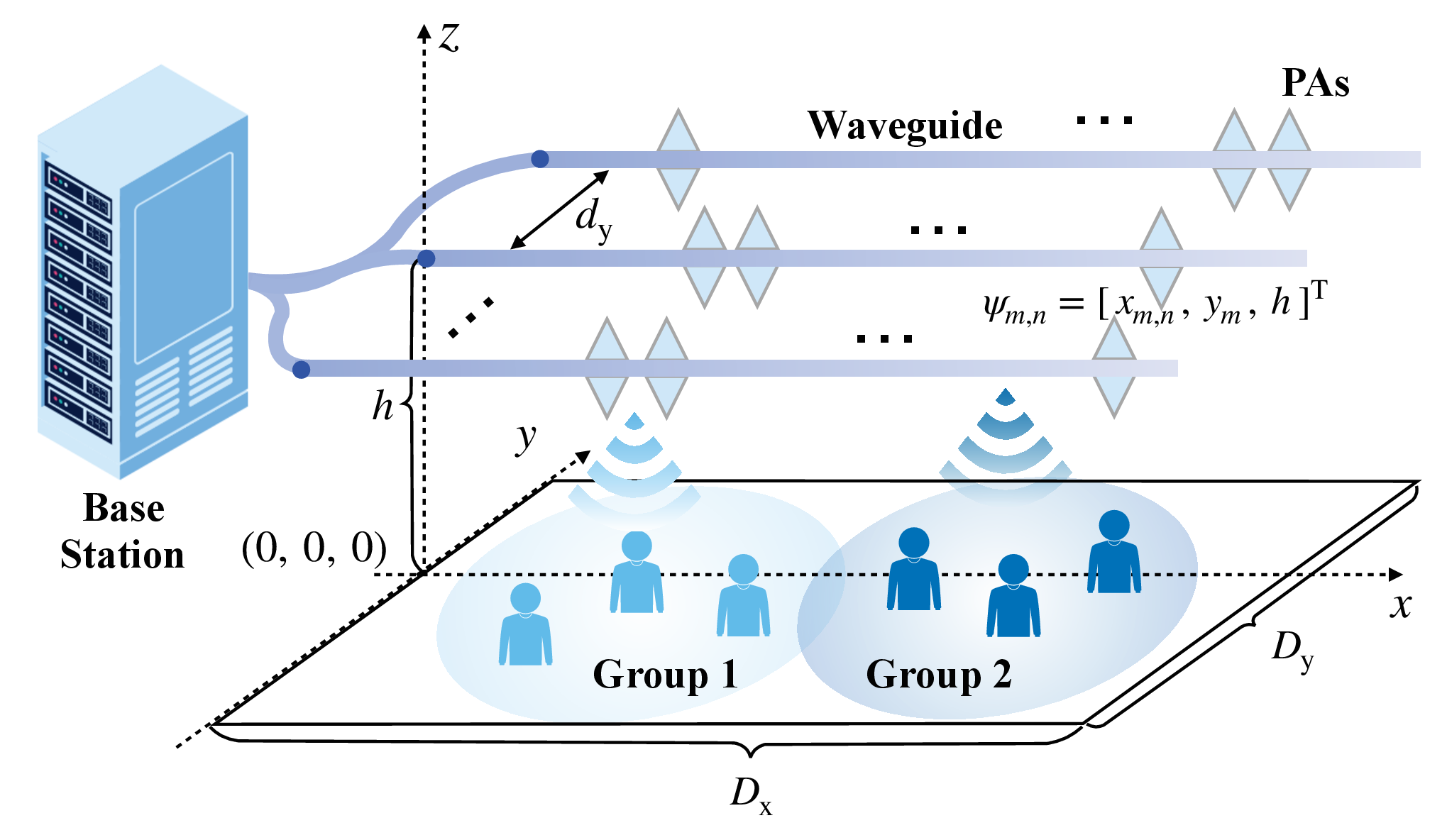}
\caption{Illustration of the PASS-based multigroup multicast system.}
\label{Fig_1}
\end{figure}

As illustrated in {\figurename}~{\ref{Fig_1}}, we consider a multigroup multicast communication system enabled by PASS. In this system, the base station (BS) employs $M$ waveguides, each activated by $N$ PAs, to serve $G$ multicast groups. Moreover, $T$ radio frequency (RF) chains are employed, which convert the signal at the baseband to high frequency, and feed it into the waveguide. To ensure that both multiplexing and division can be effectively realized, we further assume $M = T\geq G$. 

Both the waveguides and the PAs are installed at a fixed height of $h$. The PASS spans a rectangular area of size $D = D_{\mathrm{x}} \times D_{\mathrm{y}}$, and the waveguides are uniformly spaced along the $y$-axis with an equal interval of $d_{\rm y} = D_{\rm y}/(M-1)$. The Cartesian coordinate of the feed point of the $m$th waveguide is denoted as ${\psi}_{m,0} = \left[0, y_{m},  h\right]^{\rm T}$, and the position of the $n$th PA on the $m$th waveguide is given by ${\psi}_{m,n} = \left[x_{m,n},  y_{m},  h\right]^{\rm T}$, where $y_{m}$ is the $y$-coordinate of the $m$th waveguide and $x_{m,n}\in {\mathcal S}_x$ is the $x$-coordinate of the PAs with ${\mathcal S}_x$ denoting the candidate set of the PA locations along the $x$-axis. Let ${\mathcal M}\triangleq\{1,\ldots,M\}$ and ${\mathcal N}\triangleq\{1,\ldots,N\}$ denote the index sets of all waveguides and PAs per waveguide, respectively. We denote the $x$-axis positions of the PAs on the $m$th waveguide as $\mathbf{x}_{n}\triangleq\left[x_{m,1}, x_{m,2}, \dots, x_{m,N}\right]^{\rm T} \in \mathbb{R}^{N \times 1}$, which satisfy $0 \leqslant x_{n,1} < \dots < x_{n,L} \leqslant D_{\mathrm{x}}$, $ n \in \mathcal{N}$. In addition, we enforce a minimum inter-PA spacing $|x_{m,n} - x_{m,n-1}|\geq \Delta_{\rm min} = \lambda / 2$, for $m\in {\mathcal M}$, $n\in {\mathcal N}$ and $n\neq 1$, to eliminate electromagnetic mutual coupling~\cite{ouyang2025array}, where $\lambda$ is the free-space wavelength. The $x$-coordinates of all PAs across all waveguides are represented as the matrix $\mathbf{X}\triangleq\left[\mathbf{x}_{1},\mathbf{x}_{2},\dots,\mathbf{x}_{M}\right]\in\mathbb{R}^{M\times N}$.

In the multigroup multicast communication system, users within the same group receive a common data stream, while users across different groups are served with independent information. Let the set of multicast groups be denoted as $\mathcal{G}\triangleq\{1,2,\dots,G\}$. Assuming a total of $K$ users, the subset of users belonging to the group $g \in \mathcal{G}$ is denoted by $\mathcal{K}_{g}$, where each user is associated with exactly one group. That is, $\mathcal{K}_{i} \cap \mathcal{K}_{j}=\emptyset, \forall i,j \in \mathcal{G},\ i \neq j$. The coordinate of the $k$th user in the $g$th group, denoted as the $(g,k)$th user, is given by ${\hat {\bm\psi}}_{g,k} = \left[{\hat x}_{g,k},  {\hat y}_{g,k}, 0\right]^{\rm T}$.

\begin{figure*}[!t]
\centering
\includegraphics[height=0.22\textwidth]{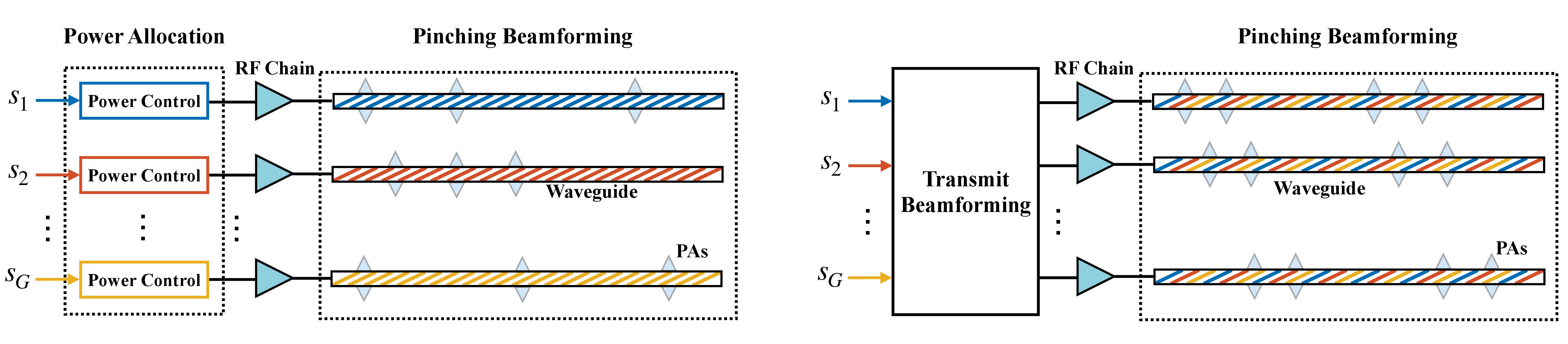}
\caption{Illustration of PASS-based multicast transmission architectures: the WD architecture (left) and the WM architecture (right).}
\label{Fig_2}
\end{figure*}

\subsection{Signal Model}
The original signal is preprocessed at the baseband through the transmit beamforming, and then converted by the RF chain and fed into the dedicated waveguide for radiation. As a result, the transmitted signal radiated by the all waveguides to the free-space can be expressed as follows:
\begin{equation}
{\mathbf{s}}=\mathbf{\Psi}(\mathbf{X})\sum\nolimits_{g=1}^{G}\mathbf{w}_{g}s_{g},
\end{equation}
where $s_{g}\in {\mathbb C}$ denotes the independent Gaussian data symbol intended for group $g$ with $\mathbb{E}\{|s_{g}|^{2}\}=1$, and $\mathbf{w}_{g}\in\mathbb{C}^{M\times1}$ represents the transmit beamforming vector of the $g$th group. Moreover, $\mathbf{\Psi}\left(\mathbf{X}\right)\in\mathbb{C}^{MN\times M}$ denotes the in-waveguide channel response when the signals are propagated from the feed point of each waveguide to specific PAs.  Since each waveguide is connected with a dedicated set of PAs, $\mathbf{\Psi}(\mathbf{X})$ can be modeled as follows:
\begin{subequations}
\begin{align}
\mathbf{\Psi}(\mathbf{X}) & = \mathrm{BlkDiag}([\boldsymbol{\psi}(\mathbf{x}_{1});\boldsymbol{\psi}(\mathbf{x}_{2});\dots;\boldsymbol{\psi}(\mathbf{x}_{M})]) \\
& =         \left[\begin{matrix}
            \boldsymbol{\psi}\left(\mathbf{x}_{1}\right) & \mathbf{0} & \dots & \mathbf{0}\\
            \mathbf{0} & \boldsymbol{\psi}\left(\mathbf{x}_{2}\right) & \dots & \mathbf{0}\\
            \vdots & \vdots & \ddots & \vdots\\
            \mathbf{0} & \mathbf{0} & \dots & \boldsymbol{\psi}\left(\mathbf{x}_{M}\right)
            \end{matrix}\right],
\end{align}
\end{subequations}
where $\boldsymbol{\psi}(\mathbf{x}_{m})\in\mathbb{C}^{N\times1}$ represents the in-waveguide propagation vector from ${\psi}_{m, 0}$ to its associated PAs. Specifically, the $n$th element of $\boldsymbol{\psi}(\mathbf{x}_{m})$ is given by:
\begin{equation}
[\boldsymbol{\psi}(\mathbf{x}_{m})]_{n}=\sqrt{P_{m,n}}{\rm exp}({-{\rm j}k_{\rm g}x_{m,n}}),
\end{equation}
where $k_{\rm g} = {2\pi}/{\lambda_{\rm g}}$ represents the guided wavenumber, and ${\lambda_{\rm g} = \lambda/n_{\rm eff}}$ is the guided wavelength with $n_{\rm eff}$ representing the waveguide's effective refractive index~\cite{10945421}.

\subsection{Transmission Model}
\subsubsection{\textbf {Power Radiation Model}}
In PASS, the power radiated at each PA is determined by two factors: the in-waveguide propagation loss and the power radiation coefficient. Let $P_m$ denote the total power allocated to the $m$th waveguide. For the $n$th PA located on the $m$th waveguide, let $\{P_{m,1}, P_{m,2}, \dots, P_{m,n-1}\}$ denote the powers radiated by each preceding PA. Subsequently, the power radiated by the $n$th PA can be modeled as follows:
\begin{align}
	P_{m,n} = \left(P_m - \sum\nolimits_{j = 1}^{n-1}P_{m,j}\right)\kappa_{m,n}a_{m,n}.
\end{align}
Here, the signal power transmission from the $(n-1)$th PA to the $n$th PA experiences:
    \paragraph{ \textbf{In-Waveguide Propagation Loss}} Expressed by:
    \begin{align}
    	\kappa_{m,n} = 10^{-\frac{\varepsilon|x_{m,n} - x_{m,n-1}|}{10}},
    \end{align}
    where $\varepsilon$ is the average attenuation factor along the dielectric waveguide in dB/m~\cite{yeh2008essence}, and $|x_{m,n} - x_{m,n-1}|$ represents the distance between adjacent PAs.
    
    \paragraph{ \textbf{Power Radiation Coefficient}} Denoted by $a_{m,n}$, representing the fraction of power radiated from the remaining waveguide power at the location of the $n$th PA. In this work, we consider two distinct power allocation models for the power radiation coefficient~\cite{wang2025modeling}:
    \begin{itemize}
    \item \textbf{Equal Power Model}: In this model, the power radiated by each PA is equal, i.e., $P_{m,n} = {P_{{\rm L},m}}/{N}$, for $m\in {\mathcal M}$ and $n\in {\mathcal N}$. Given a fixed total radiated power $P_{{\rm L},m}$, which satisfies $P_{{\rm{L}},m} \leq P_m$, the power radiation coefficient $a_{m,n}$ can be written as follows:
    \begin{align}
    a_{m,n} = \frac{P_{{\rm L}, m}}{\left(NP_{m} - (n-1)P_{{\rm L}, m}\right)\kappa_{m,n}}.
    \end{align}
    \item \textbf{Proportional Power Model}: In this model, the power radiation coefficient for each PA is equal, i.e., $a_{m,n} = a_{m}$, for $n\in {\mathcal N}$. Consequently, the radiated power at the $n$th PA follows a proportionally decaying distribution expressed as follows:
    \begin{align}
    P_{m,n} = P_{m} \left(\prod\nolimits_{j=1}^{n-1}(1 - \kappa_{m,j} a_m )\right) \kappa_{m,n} a_m .
    \end{align}
    Under this model, while $a_{m,n}$ is uniform across all PAs, the individual values $P_{m,n}$ must be determined numerically to satisfy the total power constraint.
\end{itemize}
\subsubsection{\textbf{Transmission Architecture}} 
Collecting all transmit beamforming vectors, we obtain the transmit beamforming matrix $\mathbf{W}\triangleq\left[\mathbf{w}_{1},\mathbf{w}_{2},\dots,\mathbf{w}_{G}\right]\in\mathbb{C}^{M\times G}$, subject to the power constraint $\mathrm{Tr}(\mathbf{W}^{\mathrm{H}}\mathbf{W})\leq P_{\mathrm{t}}$ with $P_{\mathrm{t}}$ denoting the power budget at the BS. For a thorough study, we consider two typical transmission architectures of PASS as illustrated in {\figurename}~{\ref{Fig_2}}~\cite{liu2025pass}:
\begin{itemize}
	\item {\bf WD}: Each waveguide delivers one multicast data stream and serves its target user group through pinching beamforming. This architecture imposes $M = G$. Therefore, the transmit beamformer can be expressed as a diagonal matrix ${\bf W} = {\rm Diag}([w_1, \cdots, w_G])$, where each $w_g$ denotes the transmit power allocated to the $g$th multicast group.
\item {\bf WM}: Each waveguide transmits a linear combination of all data streams to support spatial multiplexing, where each column vector $\mathbf{w}_g$ characterizes how the $g$th group data stream is distributed across all waveguides. Moreover, this architecture imposes $M\geq G$, which implies that more waveguides can be deployed for multiplexing. For comparison, we set $M=G$.
\end{itemize}

In summary, WD offers low implementation complexity by avoiding spatial multiplexing. However, the absence of transmit beamforming makes it less effective in managing inter-group interference. In contrast, WM enables joint transmit and pinching beamforming, which significantly improves spatial multiplexing gain, but it also introduces higher computational and hardware complexity. 

\subsection{Channel Model}
Since PASS is a promising technology in high-frequency bands~\cite{fukuda2022pinching}, where LoS propagation is generally dominant, we adopt a free-space LoS channel model. The channel vector from all the PAs to the $(g,k)$th user can be written as follows:
\begin{equation}
\mathbf{h}_{g,k}(\mathbf{X})=\left[\mathbf{h}_{g,k}^{\rm T}(\mathbf{x}_{1}),\dots,\mathbf{h}_{g,k}^{\rm T}(\mathbf{x}_{M})\right]^{\rm T}\in{\mathbb{C}}^{MN\times1},
\end{equation}
where $\mathbf{h}_{g,k}(\mathbf{x}_{m}) \in {\mathbb C}^{N\times 1}$ represents the subchannel between the user and the $m$th waveguide with
\begin{equation}
\left[\mathbf{h}_{g,k}(\mathbf{x}_{m})\right]_{n}=\frac{\sqrt{\eta}\exp\left(-{\rm j}k_0\|\boldsymbol{\psi}_{m,n}-\hat{\boldsymbol{\psi}}_{g,k}\|\right)}{\|\boldsymbol{\psi}_{m,n}-\hat{\boldsymbol{ {\psi}}}_{g,k}\|}.
\end{equation}
Here, $k_0=2\pi/\lambda$ is the free-space wavenumber, $\eta=c^2/(16\pi^2 f_{\rm c}^2)$ with $c$ and $f_{\rm c}$ representing the light speed and carrier frequency, respectively. Consequently, the received signal at the $(g,k)$th user can be expressed as follows:
\begin{align}\label{Received_Signal}
y_{g,k} & = \mathbf{h}_{g,k}^{\rm T}\left(\mathbf{X}\right)\mathbf{\Psi}\left(\mathbf{X}\right)\mathbf{w}_{g}{s}_{g} \notag \\
& +\sum\nolimits_{j\neq g}^{G} \mathbf{h}_{j,k}^{\rm T}\left(\mathbf{X}\right)\mathbf{\Psi}\left(\mathbf{X}\right)\mathbf{w}_{j}{s}_{j}+n_{g,k},
\end{align}
where $n_{g,k}\sim\mathcal{CN}(0,\sigma_{g,k}^2)$ denotes the additive Gaussian noise with $\sigma_{g,k}^2$ being the noise power. 

\subsection{Problem Formulation}
According to \eqref{Received_Signal}, the signal-to-interference-plus-noise ratio (SINR) at the $(g,k)$th user can be written as follows:
\begin{subequations}\label{eq:R_gk}
\begin{align}
 {\rm{SINR}}_{g,k}\left(\mathbf{W},\mathbf{X}\right)&=\frac{|\mathbf{\hat h}_{g,k}^{\rm T}(\mathbf{X})\mathbf{w}_{g}|^{2}}{\sum_{j\neq g}^{G}|\mathbf{\hat h}_{j,k}^{\rm T}(\mathbf{X})\mathbf{w}_{j}|^{2}+\sigma_{g,k}^{2}} \\
& \triangleq \frac{|A_{g,k}({\bf W}, {\bf X})|^{2}}{I_{g,k}({\bf W}, {\bf X})+\sigma_{g,k}^{2}},
\end{align}
\end{subequations}
where $A_{g,k}({\bf W}, {\bf X}) \triangleq \mathbf{\hat h}_{g,k}^{\rm T}(\mathbf{X})\mathbf{w}_{g}$ represents the desired signal amplitude, and $I_{g,k}({\bf W}, {\bf X})\triangleq \sum_{j\neq g}^{G}|\mathbf{\hat h}_{j,k}^{\rm T}(\mathbf{X})\mathbf{w}_{j}|^{2}$ is the inter-group interference. For simplicity, we further define 
\begin{equation} 
\mathbf{\hat h}_{g,k}^{\rm T}(\mathbf{X})\triangleq \mathbf{ h}_{g,k}^{\rm T}(\mathbf{X})\mathbf{\Psi}(\mathbf{X})\in\mathbb{C}^{1\times M} 
\end{equation} 
as the effective channel vector of the $(g,k)$th user, which is determined solely by the PA positions $\mathbf{X}$.

Due to the multicast transmission mechanism, the achievable rate of each multicast group is limited by the minimum user rate to ensure rate fairness and robust group performance, and thus the achievable multicast rate for the $g$th group is represented by
\begin{equation}\label{eq:min_rate} 
R_{g} \left(\mathbf{W},\mathbf{X}\right)=\min_{k\in\mathcal{K}{g}}\left\{ R_{g,k}\left(\mathbf{W},\mathbf{X}\right)\right\}, 
\end{equation}
where $R_{g,k}\left(\mathbf{W},\mathbf{X}\right)\triangleq\log_2(1+ {\rm{SINR}}_{g,k})$.

In this paper, we aim to jointly optimize the transmit beamformer $\mathbf{W}$ and the pinching beamformer $\mathbf{X}$ to maximize the sum rate of the multicast system. The optimization problem is thus formulated as follows:
\begin{subequations}\label{eq:Problem-formulation} 
\begin{align} 
{\mathcal P}_0: \mathop{\max}\limits_{\mathbf{W},\mathbf{X}} \  & F\left(\mathbf{W},\mathbf{X}\right)=\sum\nolimits_{g=1}^{G}R_{g} \left(\mathbf{W},\mathbf{X}\right) \\
 {\rm s.t.} & \  \mathrm{Tr}\left(\mathbf{W}^{\mathrm{\rm H}}\mathbf{W}\right)\leq P_{\mathrm{t}},\label{P_t}\\
& \ \ {x}_{m,n}\in\mathcal{S}_{x},\ \ \forall n,m,\label{S_x} \\
& \ \ |x_{m,n+1} -  x_{m,n}| \geq \Delta_{\rm min},\ \forall n,m.\label{MC}
\end{align} 
\end{subequations}

Problem ${\mathcal P}_0$ poses significant challenges due to its inherent non-convexity. Specifically, the objective function $F(\mathbf{W},\mathbf{X})$ is non-concave and non-smooth, which results from the min-rate operation across users within each group. The tight coupling between $\mathbf{W}$ and $\mathbf{X}$ further complicates the optimization. In the following sections, we develop efficient AO algorithms to address this challenging joint optimization problem.

\section{Joint Power Allocation and Pinching Beamforming Design for WD}\label{WD_architecture}
In this section, we focus on the beamforming design for WD. In this case, the transmit beamforming simplifies to the power allocation among different waveguides or groups.
\subsection{Transmit Power Allocation Optimization}
In this subsection, we focus on optimizing the transmit power allocation matrix $\mathbf{W}$ under the assumption that the PA positions $\mathbf{X}$ are fixed. Each diagonal entry $w_g$ in ${\bf W}$ can be expressed as $w_g = \sqrt{P_g}{\rm exp}({-{\rm j}\varphi_g})$ with  $P_g\in\mathbb{R}_{+}$ corresponding to the power allocated to the $g$th group, $g\in {\mathcal G}$. Let ${\bf p} = \left[{P_1}, {P_2}, \cdots, {P_G}\right]^{\rm T}\in \mathbb{R}_{+}^{G\times 1}$ denote the power allocation vector.
Consequently, the optimization of $\mathbf{W}$ simplifies to a power allocation problem subject to the total power constraint, which can be formulated as follows:
\begin{subequations}\vspace{3pt}
\begin{align} \label{eq:subProblem-F-WD} 
{\mathcal P}_1: & \mathop{\max}\limits_{{\bf p}} \ F\left(\mathbf{W}({\bf p})\right)=\sum\nolimits_{g=1}^{G}R_{g} \left({\bf p}\right) \\
 {\rm s.t.} & \  {\lVert{\mathbf{p}}\rVert^2\leq P_{\mathrm{t}}}, \ P_g\geq0, \ \forall g. \label{power-const-subProblem-F-WD}
\end{align}
\end{subequations}

The objective function~\eqref{eq:subProblem-F-WD} is non-differentiable due to the minimization operation over per-user rate. To address this issue, we employ a smooth approximation based on the following log-sum-exp (LSE) function~\cite{xu2001smoothing}:
\begin{align}\label{smoothing}
J^\tau({\bf p}) = -\frac{1}{\tau}\sum\nolimits_{g=1}^{G}\ln\left(\sum\nolimits_{k \in \mathcal{K}_g}{\rm exp}({-\tau R_{g,k}({\bf p})})\right),
\end{align}
where $\tau>0$ controls the smoothness of the approximation. The per-user rate $R_{g,k}({\bf p})$ is given by
\begin{align}\vspace{3pt}
R_{g,k}({\bf p}) = \log_2\left(1 + \frac{|\hat{h}_{g,k,(g)}|^2 \ P_g}{\sum_{j\ne g}|\hat{h}_{g,k,(j)}|^2 \ P_j + \sigma_{g,k}^2}\right),
\end{align}
with simplified expression of the effective channel gain as follows:
\begin{align}
	\hat{h}_{g,k,(g)}=[{\hat {\bf h}}_{g,k}({\bf X})]_{g}, \quad \hat{h}_{g,k,(j)}=[{\hat {\bf h}}_{g,k}({\bf X})]_{j}.
\end{align}\vspace{3pt}

To solve this approximated smooth optimization problem efficiently, we utilize the projected gradient descent (PGD) method. The gradient of $J^\tau({\bf p})$ with respect to (w.r.t.) $P_i$ is derived explicitly by applying the chain rule:
\begin{align}\label{partialrate}
\frac{\partial J^\tau({\bf p})}{\partial P_i} = \sum_{g = 1}^{G}\sum_{k\in\mathcal{K}_g}w_{g,k}\frac{\partial R_{g,k}({\bf p})}{\partial P_i},\quad i\in {\mathcal G},
\end{align}
where the weighting factor $w_{g,k}$ is computed as follows:
\begin{align}\label{lsepgd_w}
w_{g,k}=\frac{{\rm exp}\left({-\tau R_{g,k}({\bf p})}\right)}{\sum_{\ell\in\mathcal{K}_g}{\rm exp}\left({-\tau R_{g,\ell}({\bf p})}\right)}.
\end{align}
The partial derivatives of the per-user rate $R_{g,k}({\bf p})$ w.r.t the power allocation variables $P_i$ are given by
\begin{align}
\frac{\partial R_{g,k}}{\partial P_g} &= \frac{ |\hat{h}_{g,k,(g)}|^2}{\ln2\left(1+\mathrm{SINR}_{g,k}\right)I'_{g,k}({\bf p})},\label{partialrate_1} \  \ \quad i = g, \\
\frac{\partial R_{g,k}}{\partial P_i} &= -\frac{|\hat{h}_{g,k,(g)}|^2 \ |\hat{h}_{g,k,(j)}|^2 \ P_g}{\ln2\left(1+\mathrm{SINR}_{g,k}\right)I_{g,k}^{'2}({\bf p})}, \label{partialrate_2}\quad i\ne g.
\end{align}
Here, $I'_{g,k}({\bf p})$ represents the noise and the interference from other groups to the $g$th group, which reads
\begin{align}
	I'_{g,k}({\bf p}) = \sum\nolimits_{j\ne g} |\hat{h}_{g,k,(j)}|^2 \ P_j+\sigma_{g,k}^2,
\end{align}
and the SINR is given by\vspace{-5pt}
\begin{align}
	\mathrm{SINR}_{g,k}=\frac{|\hat{h}_{g,k,(g)}|^2 \ P_g}{I'_{g,k}({\bf p})}.
\end{align}\vspace{-7pt}

Substituting~\eqref{lsepgd_w}, ~\eqref{partialrate_1} and~\eqref{partialrate_2} into~\eqref{partialrate}, the iterative PGD update step is expressed as follows:
\begin{align}\label{final_p}
{\bf p}^{(t+1)} = \operatornamewithlimits{\Pi}_{ \eqref{power-const-subProblem-F-WD}}
\Bigl({\bf p}^{(t)} + \alpha^{(t)}\nabla_{\bf p} J^\tau({\bf p}^{(t)})\Bigr),
\end{align}
where $\alpha^{(t)}$ is the step size at the $t$th iteration, and $\Pi_{\eqref{power-const-subProblem-F-WD}}$ denotes projection onto the feasible set defined by the total power constraint and non-negativity constraints. This PGD strategy update iteratively refines the power allocation vector $\mathbf{p}$ within the feasible set until convergence.

\subsection{Element-Wise Sequential Optimization for Pinching Beamforming}
We now design the pinching beamforming for WD. 
Under fixed power allocation ${\bf p}$, the multigroup multicast optimization in~\eqref{eq:Problem-formulation}  reduces to
\begin{align}
	& {\mathcal P_2}: \max_{{\bf X}} \ \sum_{g = 1}^{G}\ \min_{k\in{\mathcal K}_g} \  \mathrm{SINR}_{g,k}(\mathbf{X}), \  {\rm s.t.} \, ~\eqref{S_x}, ~\eqref{MC}, \label{eq:subProblem-X-WD}
\end{align}
where
\vspace{-5pt}\begin{align}
	 \mathrm{SINR}_{g,k}(\mathbf{X}) = \frac{\left|{\bf h}_{g,k}^{\rm T}({\bf x}_{g}){\boldsymbol{\psi}}({\bf x}_g)\right|^2P_g}{\sum_{j\neq g}\left|{\bf h}_{g,k}^{\rm T}({\bf x}_{j}){\boldsymbol{\psi}}({\bf x}_j)\right|^2P_j+ \sigma_{g,k}^2}
\end{align}
denotes the SINR at the  $(g,k)$th user as a function of PA locations. 

An exhaustive search over the $ML^N$ possible configurations is computationally prohibitive for even moderate values of $N$. To reduce the computational complexity and enhance the practical feasibility of PASS, we propose an element-wise sequential optimization method, in which each $x_{m,n}\in{\bf X}$ is iteratively updated while keeping the other elements fixed. 

Specifically, problem~\eqref{eq:subProblem-X-WD} can be rewritten as follows:
\begin{align}\label{R_xmn_WD}
	\max_{x_{m,n}}\Biggl(
	    \min_{k\in{\mathcal K}_g}\mathrm{SINR}_{m,k}(x_{m,n}) +
	    \sum^{G}_{\substack{
	        {i \neq m}
	    }}
	    \min_{k\in{\mathcal K}_i} \mathrm{SINR}_{i,k}(x_{m,n})
	\Biggr),
\end{align}
To facilitate the element-wise update, we further rewrite the objective to isolate the dependency on $x_{m,n}$ as follows:
\begin{subequations}\label{final_X_WD}
\begin{align}
	& {\mathcal P_3}: 	\max_{x_{m,n}}\Biggl(
	    \min_{k\in{\mathcal K}_g}\frac{A_{k}(x_{m,n})}{C_{m,k}^0} +
	    \sum^{G}_{\substack{
	        {i \neq m}
	    }}
	    \min_{k\in{\mathcal K}_i} \frac{C_{i,k}^1}{A_{k}(x_{m,n})+C_{i,k}^2}
	\Biggr),\label{x_mn_WD} \\
	& {\rm s.t.} \, ~\eqref{S_x}, ~\eqref{MC},
\end{align}
\end{subequations}
where $A_{k}(x_{m,n}) = \left|{\bf h}_{m,k}^{\rm T}({\bf x}_{m}){\boldsymbol{\psi}}({\bf x}_m)\right|^2P_m$ represents the residual term that depends on $x_{m,n}$, and the constant terms independent of $x_{m,n}$ can be precomputed as follows:
\begin{subequations}
\begin{align}
	& C_{i,k}^1 = \left|{\bf h}_{i,k}^{\rm T}({\bf x}_{i}){\boldsymbol{\psi}}({\bf x}_i)\right|^2P_i, \\
	& C_{m,k}^0 = \frac{\sum_{i\neq m}^{G}C_{i,k}^1 + \sigma_{m,k}^2}{P_m}, \\
	& C_{i,k}^2 = \sum\nolimits_{{j\neq i, {j \neq m}}}^{G} C_{j,k}^1 + \sigma_{i,k}^2.
\end{align}
\end{subequations}
In addition, the feasible set $\mathcal{S}_{x}$ in \eqref{S_x} is discretized into a uniform grid of candidate locations to facilitate one-dimensional search, which can be expressed as follows:
\begin{equation}\label{Index}
\mathcal{S}_{x} = \left\{\,0, \frac{D_{\rm x}}{L-1},\, \frac{2D_{\rm x}}{L-1},\, \dots, \,D_{\rm x}\,\right\},
\end{equation} 
where $L = |\mathcal S_x|$ denotes the number of discrete positions available for deploying the PAs.

Problem~\eqref{x_mn_WD} reduces to optimizing a single variable over a bounded interval. By calculating and fixing the constant terms $\left\{C_{m,k}^0, C_{i,k}^1, C_{i,k}^2\right\}_{k = 1}^{K}$ beforehand, the optimal $x_{m,n}$ can be efficiently obtained via a low-complexity one-dimensional grid search over the feasible set ${\mathcal S}_x$.

\begin{algorithm}[t]
\caption{AO Algorithm for Solving~\eqref{eq:Problem-formulation} in the WD Architecture}
\label{algo:AO}
\begin{algorithmic}[1]
\STATE initialize $\mathbf{p}^{(0)},\mathbf{X}^{(0)}$, set threshold $\epsilon$, $t\gets0$
\REPEAT
    \STATE update $\mathbf{p}^{(t+1)}$ by solving~\eqref{final_p}
    \STATE update each $x_{m,n}^{(t+1)}$ by solving~\eqref{final_X_WD}
    \STATE $t\gets t+1$
\UNTIL $\|\mathbf{p}^{(t)}-\mathbf{p}^{(t-1)}\|\le\epsilon$ and $\|\mathbf{X}^{(t)}-\mathbf{X}^{(t-1)}\|\le\epsilon$
\end{algorithmic}
\end{algorithm}

\subsection{Complexity and Convergence Analysis}
The proposed AO algorithm for solving~\eqref{eq:Problem-formulation} in the WD architecture is summarized in Algorithm~\ref{algo:AO}.
\subsubsection{\bf Complexity Analysis}
The complexity of the PGD approach for solving subproblem $\mathcal{P}_1$ is $\mathcal{O}(I_1GK)$, where $I_1$ denotes the number of PGD iterations.
For subproblem $\mathcal{P}_3$, the element-wise sequential update incurs a complexity of $\mathcal{O}(GKMNL)$.
As a result, assuming $I_2$ iterations in the AO framework, the total computational complexity of Algorithm~\ref{algo:AO} is given by $\mathcal{O}\bigl(I_2(I_1GK + GKMNL)\bigr)$.

\subsubsection{\bf Convergence Analysis}
At each AO iteration, the PGD update for $\mathbf F$ in $\mathcal P_1$ and the element-wise sequential search for $\mathbf X$ in $\mathcal P_2$ both yield non-decreasing values of the original objective in~\eqref{eq:Problem-formulation}.  Since this objective is limited by the total transmit power constraint and the finite cardinality of $\mathcal S_x$, the sequence of objective values generated by Algorithm~\ref{algo:AO} is monotonic and bounded, and thus converges to a suboptimal point.  

\section{Joint Transmit Beamforming and Pinching Beamforming Design for WM}\label{WM_architecture}
In this section, we study the joint transmit beamforming and pinching beamforming design under the WM architecture.

\subsection{MM-based Reformulation}
For WM, the original problem in~\eqref{eq:Problem-formulation} is a highly non-convex fractional-quadratic program. In particular, even with ${\bf X}$ fixed, the resulting subproblem in ${\bf W}$ remains non-convex and non-smooth. Therefore, directly using the AO framework or gradient-based methods cannot guarantee convergence. To handle this difficulty, we adopt the MM technique to transform the original objective function into a more tractable surrogate problem, in which the AO framework can be developed to update the transmit and pinching beamformer iteratively.

Specifically, let $\{{\bf W}^{(t)},{\bf X}^{(t)}\}$ denote the solutions obtained from the previous iteration $t-1$. Then, $R_{gk}\left({\bf W},{\bf X}\right)$ is minorized by a concave surrogate function $\widetilde{R}_{gk}\left({\bf W},{\bf X}|{\bf W}^{(t)},{\bf X}^{(t)}\right)$, defined as follows:
\begin{align}\label{eq:Rate-surrogate}
& \widetilde{R}_{g,k}\left({\bf W},{\bf X}|{\bf W}^{(t)},\mathbf{X}^{(t)}\right)\notag \\
&={\rm const}_{g,k}+2\Re\left\{ a_{g,k}A_{g,k}({\bf W}, {\bf X})\right\}\notag\\
 & -b_{g,k}\left(I_{g,k}({\bf W}, {\bf X}) + |A_{g,k}({\bf W}, {\bf X})|^{2}\right)\nonumber \\
 & \leq R_{g,k}\left({\bf W},{\bf X}\right),
\end{align}
where
\begin{subequations}\label{surrogate_total}
\begin{align}
 & a_{g,k}=\frac{A_{g,k}^{\rm H}({\bf W}^{(t)}, {\bf X}^{(t)})}{I_{g,k}({\bf W}^{(t)}, {\bf X}^{(t)})+\sigma_{g,k}^{2}},\label{Rate-surrogate-a}\\
 & b_{g,k}=\frac{|A_{g,k}({\bf W}^{(t)}, {\bf X}^{(t)})|^{2}\left(I_{g,k}({\bf W}^{(t)}, {\bf X}^{(t)})+\sigma_{g,k}^{2}\right)^{-1}}{\left(I_{g,k}({\bf W}^{(t)}, {\bf X}^{(t)})+|A_{g,k}({\bf W}^{(t)}, {\bf X}^{(t)})|^{2}+\sigma_{g,k}^{2}\right)},\label{Rate-surrogate-b}\\
 & {\rm const}_{g,k}=R_{g,k}\left({\bf W}^{(t)},{\bf X}^{(t)}\right)-2 b_{g,k}\sigma_{g,k}^{2}\label{Rate-surrogate-const}\nonumber\\
 & \ \ \ \ \ \ \ \ \  -b_{g,k}\left(I_{g,k}({\bf W}^{(t)}, {\bf X}^{(t)})+|A_{g,k}({\bf W}^{(t)}, {\bf X}^{(t)})|^{2}\right),
\end{align}
\end{subequations}
at the fixed point $\{\mathbf{W}^{(t)},\mathbf{X}^{(t)}\}$.
The proof of~\eqref{eq:Rate-surrogate} is similar to the one provided in \cite{9076830}, which employs first-order Taylor expansion to construct a suitable surrogate function. The detailed derivation is omitted here for brevity.

Based on the MM surrogate, the original optimization problem (\ref{eq:Problem-formulation}) can be equivalently approximated by the following surrogate problem at iteration $t$:
\begin{subequations}\label{eq:Problem-MM}
\begin{align}
{\mathcal P_4:}\mathop{\max}\limits _{\mathbf{W},\mathbf{X}}&\quad \widetilde{F}\left(\mathbf{W},\mathbf{X}|\mathbf{W}^{(t)},\mathbf{X}^{(t)}\right)\nonumber\\
&=\sum\nolimits_{g=1}^{G}\min_{k\in\mathcal{K}_{g}}\left\{ \widetilde{R}_{g,k}\left(\mathbf{W},\mathbf{X}|\mathbf{W}^{(t)},\mathbf{X}^{(t)}\right)\right\}\\
& {\rm s.t.}\quad \eqref{P_t},\, \eqref{S_x},\, \eqref{MC}.
\end{align}
\end{subequations}
We note that the surrogate function $\widetilde{R}_{g,k}\left(\mathbf{W},\mathbf{X}|\mathbf{W}^{(t)},\mathbf{X}^{(t)}\right)$ defined in (\ref{eq:Rate-surrogate}) is concave in $\mathbf{W}$ for a fixed $\mathbf{X}$. This property enables an effective AO framework to solve problem (\ref{eq:Problem-MM}), where $\mathbf{W}$ and $\mathbf{X}$ can be iteratively optimized until convergence.

\subsection{Optimization of Transmit Beamforming}
In this subsection, we optimize the transmit beamformer $\mathbf{W}$ for a given pinching beamformer $\mathbf{X}$.
Here, we formulate a Lagrangian function based on the MM-approximated objective, analytically derive the optimal structure of $\mathbf{W}$, and finally optimize the transmit beamforming based on this closed-form solution. 

To efficiently solve problem \eqref{eq:Problem-MM}, we first rewrite the surrogate function $\widetilde{R}_{g,k}(\mathbf{W}|\mathbf{W}^{(t)})$ defined in \eqref{eq:Rate-surrogate} as a quadratic function w.r.t. the transmit beamformer $\mathbf{W}$, which can be expressed as follows:
\begin{align}\label{eq:surrogate-F}
\widetilde{R}_{g,k}\left(\mathbf{W}|\mathbf{W}^{(t)}\right) & =\textrm{const}_{g,k}+2\Re \left\{ a_{g,k}\mathbf{\hat h}_{g,k}^{\mathrm{T}}{\bf w}_{g}\right\} \notag \\
& -b_{g,k}\sum_{i=1}^{G}|\mathbf{\hat h}_{g,k}^{\mathrm{T}}{\bf w}_{i}|^{2}.
\vspace{-5pt}\end{align}
Consequently, the transmit beamformer optimization subproblem becomes
\vspace{-5pt}\begin{align}\label{eq:F-original}
 \max_{\mathbf{W}} \ \sum_{g=1}^{G}\min_{k\in\mathcal{K}_{g}}\widetilde{R}_{g,k}\left(\mathbf{W}|\mathbf{W}^{(t)}\right),\
 {\rm s.t.} \ \eqref{P_t}.
\end{align}

To further simplify the optimization, we introduce auxiliary variables $\boldsymbol{\gamma}=[\gamma_{1},\dots,\gamma_{G}]^{\mathrm{T}}$ to transform problem \eqref{eq:F-original} into a standard convex second-order cone program (SOCP):
\begin{subequations}\label{eq:Problem-F-socp-final}
\begin{align}
{\mathcal P}_5: &\max_{\mathbf{W},\boldsymbol{\gamma}} \quad \sum_{g=1}^{G}\gamma_{g}\\
{\rm s.t.} &\quad\widetilde{R}_{g,k}\left(\mathbf{W}|\mathbf{W}^{(t)}\right)\geq\gamma_{g},\,\, \forall k, g, \quad \eqref{P_t}.
\end{align}
\end{subequations}
At this stage, problem \eqref{eq:Problem-F-socp-final} can be readily solved by existing CVX solvers such as MOSEK~\cite{mosek2015}. However, this direct approach incurs substantial computational overhead due to the high complexity associated with canonicalizing and parsing the problem, particularly for large-scale systems.

\subsubsection{Lagrangian Function Formulation and the Optimal Transmit Beamformer Structure}
To reduce this computational complexity, we propose an efficient algorithm by leveraging the analytical structure of the optimal transmit beamformer. Specifically, we first formulate the Lagrangian function associated with problem \eqref{eq:Problem-F-socp-final} as follows:
\begin{align}\label{Lag_F}
&\mathcal{L}\Bigl({\bf W}, {\boldsymbol{\gamma}}, \{{\boldsymbol{\delta}}_g\}_{g\in{\mathcal G}}, \nu\Bigr)=\sum_{g=1}^G \gamma_g -\nu\left(\sum_{g=1}^G \|\mathbf{w}_g\|^2 - P_{\mathrm{t}}\right) \nonumber \\
&\qquad\qquad -\sum_{g=1}^G\sum_{k\in\mathcal{K}_g}\delta_{g,k}\left(\gamma_g -\widetilde{R}_{g,k}\left(\mathbf{W}|\mathbf{W}^{(t)}\right)\right),
\end{align}
where ${\boldsymbol{\delta}}_g \triangleq [\delta_{g,1}, \cdots, \delta_{g,K}]^{\rm T}$, $\delta_{g,k}\geq 0$ and $\nu\geq0$ are dual variables corresponding to the respective constraints in \eqref{eq:Problem-F-socp-final}. By setting the first-order derivative of $\mathcal{L}\left({\bf W}, {\boldsymbol{\gamma}}, \{{\boldsymbol{\delta}}_g\}_{g\in{\mathcal G}}, \nu\right)$ w.r.t. ${\bf w}_g^*$ to zero, we have
\begin{align}
	& \frac{\partial \mathcal{L}\left({\bf W}, {\boldsymbol{\gamma}}, \{{\boldsymbol{\delta}}_g\}_{g\in{\mathcal G}}, \nu\right)}{\partial {\bf w}_g^*} \notag \\
	& = \sum_{k\in\mathcal K_g}\delta_{g,k}\,
\frac{\partial\widetilde R_{g,k}\left(\mathbf{W}|\mathbf{W}^{(t)}\right)}{\partial {\bf w}_g^*}
-\nu\,{\bf w}_g = 0.
\end{align}
Recall that the partial derivative of $\widetilde{R}_{g,k}\left(\mathbf{W}|\mathbf{W}^{(t)}\right)$ w.r.t. ${\bf w}_g^*$ can be given as follows:
\begin{align}
\frac{\partial\widetilde{R}_{g,k}\left(\mathbf{W}|\mathbf{W}^{(t)}\right)}{\partial {\bf w}_g^*}
&=a_{g,k}\,{\bf \hat h}_{g,k}
-b_{g,k}\sum_{i=1}^G{\bf \hat h}_{g,k}{\bf \hat h}_{g,k}^{\rm H}\,{\bf w}_i.
\end{align}
Then the globally optimal solution of ${\bf w}_g$ in the $t$th iteration can be derived as follows:
\begin{equation}\label{eq:optimal-f}
\mathbf{w}_g^{\star}=\left(\sum_{i=1}^{G}\sum_{k\in\mathcal{K}_{i}}\delta_{i,k}b_{i,k}\mathbf{\hat{h}}_{i,k}\mathbf{\hat{h}}_{i,k}^{\mathrm{H}}+\nu\mathbf{I}\right)^{-1}\sum_{k\in\mathcal{K}_g}\delta_{g,k}a_{g,k}\mathbf{\hat{h}}_{g,k}.
\end{equation}
Additionally, setting $\partial\mathcal{L}\left({\bf W}, {\boldsymbol{\gamma}}, \{{\boldsymbol{\delta}}_g\}_{g\in{\mathcal G}}, \nu\right)/\partial\gamma_g=0$ gives
\begin{align}\label{delta_gk>0}
\sum_{k\in\mathcal K_g}\delta_{g,k}=1,\quad g\in{\mathcal G}.
\end{align}
\subsubsection{Lagrangian Dual Problem and PAGD Updates}
Substituting \eqref{eq:optimal-f} into the Lagrangian function \eqref{Lag_F}, the optimization reduces to solving the Lagrange dual problem, which can be expressed as follows:
\begin{subequations}\label{dual-problem}
\begin{align}
{\mathcal P}_6: & \min_{{\boldsymbol{\gamma}}, \{{\boldsymbol{\delta}}_g\}_{g\in{\mathcal G}}, \nu}\mathcal{D}\Bigl({\bf W},\ {\boldsymbol{\gamma}}, \{{\boldsymbol{\delta}}_g\}_{g\in{\mathcal G}},\ \nu\Bigr) \\
 {\rm s.t.}& \quad \nu\geq 0, \\
 & \quad \delta_{g,k}\in {\mathcal H}_g, \ k\in\mathcal{K}_{g},\, \forall g\in\mathcal{G}, \label{sum_delta}
\end{align}
\end{subequations}
where the Lagrange dual function is given by
\begin{align}
\mathcal{D}\Bigl({\bf W},\ {\boldsymbol{\gamma}}, \{{\boldsymbol{\delta}}_g\}_{g\in{\mathcal G}},\ \nu\Bigr)&=\sum_{g=1}^G\sum_{k\in\mathcal{K}_g}\delta_{g,k}\widetilde{R}_{g,k}\bigl(\mathbf{W}^\star|\mathbf{W}^{(t)}\bigr)\notag \\
& -\nu\left(\sum_{g = 1}^{G}\|\mathbf{w}_g^\star\|^2 - P_{\mathrm{t}}\right).
\end{align}
Furthermore, the constraint~\eqref{sum_delta} is derived from \eqref{delta_gk>0}, where the feasible space of $\delta_{g,k}$ can be written as follows:
\begin{align}
	{\mathcal H}_g \triangleq \left\{\delta_{g,k} \geq 0: \sum_{k \in {\mathcal K}_g} \delta_{g,k} = 1\right\}.
\end{align}

The dual problem in~\eqref{dual-problem} is convex, and each dual vector ${\boldsymbol{\delta}}_g$ lies on a hyperplane ${\mathcal H}_g$. This structure motivates us to solve~\eqref{dual-problem} using the PAGD algorithm~\cite{fang2023optimal}, which iteratively minimizes the objective by updating the solution in the negative gradient direction with an adaptive step size at each iteration. A subsequent projection step ensures that the updated solution remains within the feasible space.
 \paragraph{Update of $\{{\boldsymbol{\delta}}_g\}_{g\in{\mathcal G}}$}
 Specifically, the dual variables $\delta_{g,k}$ are first updated by
\begin{align}\label{iter_delta}
\bar{\delta}_{g,k}^{(q+1)}&=\delta_{g,k}^{(q)}-\mu_{g,k}^{(q)}g_{\delta_{g,k}}^{(q)},\quad g\in{\mathcal G}, k\in{\mathcal K}_g,
\end{align}
where $\bar{\delta}_{g,k}^{(q+1)}$ represents the intermediate variable updates before projection, $g_{\delta_{g,k}}^{(q)}$ represents the subgradient of $\delta_{g,k}$ in the $q$th iteration, which can be given as follows:
\begin{align}\label{subgradient_delta}
g_{\delta_{g,k}}^{(q)} =\widetilde R_{g,k}^{(q)} -\min_{\ell\in\mathcal K_g}\widetilde R_{g,\ell}^{(q)}.
\end{align}
This subgradient follows directly from the stationarity condition of the Lagrangian function w.r.t $\gamma_g$. Specifically, the derivative of the partial Lagrangian w.r.t. each $\delta_{g,k}$ can be given as follows:
\begin{align}\label{partial_L_delta}
\frac{\partial \mathcal{L}\left({\bf W}, {\boldsymbol{\gamma}}, \{{\boldsymbol{\delta}}_g\}_{g\in{\mathcal G}}, \nu\right)}{\partial \delta_{g,k}}
=-\bigl(\gamma_g-\widetilde R_{g,k}\bigr), \ g\in{\mathcal G}, k\in{\mathcal K}_g.
\end{align}
By solving the inner maximization w.r.t. $\gamma_g$, the optimal solution is given by
\begin{align}\label{optimal_gamma}
	\gamma_g^\star
=\min_{\ell\in\mathcal K_g}\widetilde R_{g,\ell},\quad g\in{\mathcal G},
\end{align}
since that is the largest $\gamma_g$ that satisfies all $\gamma_g\le\widetilde R_{g,k}$.  Substituting~\eqref{optimal_gamma} into~\eqref{partial_L_delta}, the subgradient expression in~\eqref{subgradient_delta} can be obtained.

In addition, $\mu_{g,k}^{(q)}$ in~\eqref{iter_delta} denotes an adaptive step size, which can be chosen as follows~\cite{NedicBertsekas2001}:
\begin{align}\label{update_mu}
\mu_{g,k}^{(q)}=\frac{\delta_{g,k}^{(q)}}{\widetilde{R}_{g,k}^{(q)}-\min_{\ell\in\mathcal{K}_g}\widetilde{R}_{g,\ell}^{(q)}+\rho_c+{q}\rho_{\mu}}, \quad g\in{\mathcal G}, k\in{\mathcal K}_g.
\end{align}
Here, $\rho_c+q\rho_{\mu}$ represents an increasing constant number with $\rho_c > 0$ and $\rho_{\mu} > 0$.

Subsequently,  by projecting  $\bar{\delta}_{g,k}^{(q+1)}$ onto the feasible space $\sum_{k\in {\mathcal K}_g}\delta_{g,k}=1,\delta_{g,k}\geq 0$, the dual variables $\delta_{g,k}$ can be obtained as follows~\cite{Michelot1986}:
\begin{align}\label{updata_delta}
\bm{\delta}_g^{(q+1)}=\bar{\bm{\delta}}_g^{(q+1)}-\frac{\sum_{k\in {\mathcal K}_g}\bar{\delta}_{g,k}^{(q+1)}-1}{K_g}\mathbf{1}_{G}, \quad g\in{\mathcal G},
\end{align}
where $\mathbf{1}_{G} \triangleq [1, \cdots, 1]^{\rm T}\in {\mathbb R}^{G\times 1}$.
\paragraph{Update of $\nu$}
In the Lagrangian problem~\eqref{Lag_F}, the subgradient of the dual function w.r.t. $\nu$ at the $q$th iteration can be given as follows:
\begin{align}
g_\nu^{(q)}
&= \frac{\partial}{\partial \nu} \left( - \nu \left( \sum_{g = 1}^{G} \left\Vert\mathbf w_g^{(q)}\right\Vert^2 - P_{\rm t} \right) \right)\notag \\
& = \sum_{g = 1}^{G} \left\Vert{\bf w}_g^{(q)}\right\Vert^2 - P_{\rm t}.
\end{align}
Applying the projected subgradient method, $\nu$ is first updated by
\begin{align}
\bar\nu^{(q+1)}
&= \nu^{(q)} - \xi^{(q)} \, g_\nu^{(q)} \nonumber\\
&= \nu^{(q)} - \xi^{(q)} \left( \sum_{g = 1}^{G} \left\Vert{\mathbf w}_g^{(q)}\right\Vert^2 - P_{\rm t} \right),
\end{align}
and then projected onto the nonnegative orthant, the dual variable $\nu$ can be updated as follows:
\begin{align}\label{update_tau}
\nu^{(q+1)} = \max \left\{ 0, \,\,\bar\nu^{(q+1)} \right\}.
\end{align}
The step size can be chosen as follows~\cite{NedicOzdaglar2009}:
\begin{align}\label{update_xi}
\xi^{(q)} = \frac{\rho_p}{\rho_t + {q}\, \rho_{\eta}}, \quad \rho_p,\, \rho_t,\, \rho_{\eta} > 0,
\end{align}
which satisfies the standard diminishing-step-size conditions and thus ensures convergence of the dual iterations.

Overall, by embedding the proposed PAGD approach within the MM optimization framework, we obtain a computationally efficient MM-PAGD algorithm for the transmit beamformer update.

\subsection{Optimization of Pinching Beamforming}
In this section, we focus on optimizing the pinching beamforming matrix ${\bf X}$ with given ${\bf F}$. According to the surrogate rate expression in \eqref{eq:Problem-MM}, the subproblem of optimizing the pinching beamforming can be rewritten as follows:
\begin{subequations}\label{subproblem_X}
\begin{align}
	{\mathcal P}_7: & \mathop{\max}\limits _{\mathbf{X}}\sum_{g=1}^{G}\min_{k \in {\mathcal K}_g}\widetilde{R}_{g,k}\left(\mathbf{X}|\mathbf{X}^{(t)}\right) \label{MM-X} \\
    {\rm s.t.} \,\, & \eqref{S_x},\, \eqref{MC}.
\end{align}
\end{subequations}
where
\begin{align}\label{eq:rate-X}
 \widetilde{R}_{g,k}\left(\mathbf{X}|\mathbf{X}^{(t)}\right) & = {\rm const}_{g,k}+2\Re\left\{ a_{g,k}A_{g,k}({\bf X})\right\}\notag \\
& -b_{g,k}\left(I_{g,k}({\bf X}) + |A_{g,k}({\bf X})|^{2}\right).
\end{align}
Similar to WD, we adopt an element-wise sequential approach for the pinching beamforming optimization. 

First, we derive a partial objective function by isolating the dependency on a single PA position $x_{m,n}$ based on the MM surrogate multicast rate expression. Specifically, the MM-based partial objective function of~\eqref{subproblem_X} can be expressed as follows:
\begin{subequations}
\begin{align}\label{per-group-rate}
		{\mathcal P}_{8}: & \mathop{\max}\limits _{x_{m,n}}\sum_{g=1}^{G}\min_{k \in {\mathcal K}_g}\widetilde{R}_{g,k}\left(x_{m,n}|\mathbf{X}^{(t)}\right) \\
	     {\rm s.t.} \,\, & \eqref{S_x},\, \eqref{MC},\, \eqref{Index}.
\end{align}
\end{subequations}
The per-group rate in~\eqref{per-group-rate} can be further reorganized as follows:
\begin{align}\label{R_1+R_2}
	 \widetilde{R}_{g,k}\left(x_{m,n}|\mathbf{X}^{(t)}\right) & =  \widetilde{R}^1_{g,k}\left(A_{g,k}({\bf X})\right) +\widetilde{R}^2_{g,k}\left(I_{g,k}({\bf X})\right),
\end{align}
where
\begin{align}\label{R_1}
	 \widetilde{R}^1_{g,k}\left(A_{g,k}({\bf X})\right) \triangleq -b_{g,k}|A_{g,k}({\bf X})|^{2} + 2\Re\left\{ a_{g,k}A_{g,k}({\bf X})\right\}
\end{align}
and
\begin{align}\label{R_2}
	\widetilde{R}^2_{g,k}\left(I_{g,k}({\bf X})\right) \triangleq -b_{g,k}I_{g,k}({\bf X}).
\end{align}
\vspace{-1pt}
The term $A_{g,k}({\bf X})$ in~\eqref{R_1} can be expanded w.r.t $x_{m,n}$ as follows:
\begin{align}\label{A_x_mn}
	A_{g,k}(x_{m,n}) & = S_{g,k}^{m-} + A_{g,k, (g)}({\bf x}_{m}),
\end{align}
where
\begin{align}
	A_{g,k, (g)}({\bf x}_{m}) = {\bf h}_{g,k}^{\rm T}({\bf x}_{m})\boldsymbol{\psi}({\bf x}_{m})\left[{\bf f}_g\right]_{m}
\end{align}
represents a residual term that depends on $x_{m,n}$ and
\begin{align}
	S_{g,k}^{m-} = \sum_{l\neq m}^{M}{\bf h}_{g,k}^{\rm T}({\bf x}_{l})\boldsymbol{\psi}({\bf x}_{l})\left[{\bf f}_g\right]_{l}
\end{align}
is independent of $x_{m,n}$.
Similarly, the term $I_{g,k}({\bf X})$ in~\eqref{R_2} can be expanded as follows: 
\begin{align}\label{All_x_mn}
	& I_{g,k}({\bf X}) = \sum_{j \neq g}\left(S_{g,k,(j)}^{m-} + I_{g,k,(j)}({\bf x}_{m})\right)^2,
\end{align}
where
\begin{align}
	S_{g,k,(j)}^{m-} = \sum_{l\neq m}^{M}{\bf h}_{g,k}^{\rm T}({\bf x}_{l})\boldsymbol{\psi}({\bf x}_{l})\left[{\bf f}_j\right]_{l}
\end{align}
 and
 \begin{align}
 	I_{g,k,(j)}({\bf x}_{m}) = {\bf h}_{g,k}^{\rm T}({\bf x}_{m})\boldsymbol{\psi}({\bf x}_{m})\left[{\bf f}_j\right]_{m}.
 \end{align}
 Taken together, the optimization of $x_{m,n}$ in the element-wise sequential optimization framework is formulated as follows:
\begin{subequations}\label{subproblem_propor}
\begin{align}
	{\mathcal P}_{9}: & \mathop{\max}\limits _{x_{m,n}}\sum_{g=1}^{G}\min_{k \in {\mathcal K}_g}\widetilde{R}_{g,k}\left(x_{m,n}|\mathbf{X}^{(t)}\right) \\
	{\rm s.t.}& \,\, \eqref{S_x},\, \eqref{MC},\, \eqref{Index}.
\end{align}
\end{subequations}
where the expression of $\widetilde{R}_{g,k}\left(x_{m,n}|\mathbf{X}^{(t)}\right)$ shown at the top of the next page, and 
\begin{subequations}
\begin{align}
	 C^1_{i,k} &= -b_{g,k}, \\
	 C^2_{m,k} &= a_{g,k} - \frac{1}{2}b_{g,k}S_{g,k}^{m-}, \\
	 C^3_{i,k} &= \frac{1}{2}b_{g,k}S_{g,k,(j)}^{m-}.
\end{align}
\end{subequations}

By calculating and fixing the constant terms $\left\{C^1_{i,k}, C^2_{m,k}, C^3_{i,k}\right\}_{\substack{i,m\in {\mathcal G} \\ k \in {\mathcal K}_g}}$ beforehand, the optimal $x_{m,n}$ can be efficiently obtained via a low-complexity one-dimensional grid search over the feasible set ${\mathcal S}_x$.
\begin{figure*}
	\begin{align}
		\widetilde{R}_{g,k}\left(x_{m,n}|\mathbf{X}^{(t)}\right) = C^1_{i,k}|A_{g,k}({\bf x}_{m})|^{2} + 2\Re\left\{C^2_{m,k}A_{g,k}({\bf x}_{m})\right\} + \sum_{j\neq g}C^1_{i,k}|I_{g,k,(j)}({\bf x}_{m})|^2 - \sum_{j\neq g}2\Re\left\{C^3_{i,k}I_{g,k,(j)}({\bf x}_{m})\right\}.
	\end{align}
	\hrulefill
\end{figure*}
The solution $x_{m,n}^{(t+1)}$ is then adopted for the next iteration.

\begin{algorithm}[t]
\caption{AO Algorithm for Solving~\eqref{eq:Problem-formulation} in the WM Architecture}
\label{algo:MM-PAGD-AO}
\begin{algorithmic}[1]
\STATE initialize $\mathbf W^{(0)}, \mathbf X^{(0)}, \epsilon,\, t \gets 0$

\REPEAT
    \STATE compute surrogate parameters $\{a_{g,k}, b_{g,k}, \mathrm{const}_{g,k}\}$ at $(\mathbf W^{(t)}, \mathbf X^{(t)})$ by solving~\eqref{surrogate_total}
    \STATE update $\mathbf X^{(t+1)}$ by solving~\eqref{subproblem_propor}
    \STATE update dual variables $\{\{{\boldsymbol{\delta}}_g\}_{g\in{\mathcal G}},\ \nu\}$ by solving~\eqref{updata_delta} and~\eqref{update_tau}
    \STATE compute $\mathbf w_g^{\star}$ by solving \eqref{eq:optimal-f}
    \STATE $t \gets t + 1$
\UNTIL $\|\mathbf X^{(t)} - \mathbf X^{(t-1)}\| \le \epsilon$ and $\|\mathbf W^{(t)} - \mathbf W^{(t-1)}\| \le \epsilon$
\end{algorithmic}
\end{algorithm}

\subsection{Complexity and Convergence Analysis}
The proposed AO algorithm for solving~\eqref{eq:Problem-formulation} in the WM architecture is summarized in Algorithm~\ref{algo:MM-PAGD-AO}.
\subsubsection{\bf Complexity Analysis}
The computational complexity is dominated by three components per outer AO iteration: surrogate function update with complexity $\mathcal{O}(GKM)$, transmit beamforming via PAGD with $Q$ iterations each costing $\mathcal{O}(GKM^2 + M^3 + GK + GM)$, and pinching beamforming via element-wise grid search with complexity $\mathcal{O}(MNLGK)$. As a result, the total complexity scales as $\mathcal{O}\bigl(T_3[GKM + Q(GKM^2 + M^3 + GK + GM) + MNLGK]\bigr)$, where $T_3$ is the number of AO iterations.

\subsubsection{\bf Convergence Analysis}
At each iteration, the MM surrogate guarantees that the joint update of $\mathbf F$ and $\mathbf X$ does not decrease the original objective in~\eqref{eq:Problem-formulation}. Since this objective is limited by total transmit power constraint, the sequence of objective values generated by Algorithm~\ref{algo:MM-PAGD-AO} is monotonically non-decreasing and bounded above. Therefore, the algorithm converges to a suboptimal solution of the original non-convex problem.

\section{Numerical Results}\label{simulation}
We now validate the effectiveness of the proposed PASS framework for multigroup multicast communications through comprehensive numerical simulations. Specifically, we compare the performance of PASS with conventional MIMO and massive MIMO baselines under the same simulation setup. Our evaluation focuses on several aspects: (i) the multicast rate improvements achieved by PASS relative to fixed-location antenna systems; (ii) the impact of different power radiation models on the overall system performance; and (iii) a comparative analysis of the WD and WM architectures across diverse deployment scenarios to identify their respective strengths and limitations. 

\subsection{Simulation Setup}
For the configuration of PASS, the waveguides are deployed at a height of $h = 5$ m with a total length of $D_{\rm x}$. The other side length of the rectangular region is set to $D_{\rm y} = 5$ m. The number of discrete search points in the one-dimensional search is set to $L = 10^3$. The PAs are deployed along the waveguide with a minimum separation of $\Delta_{\rm min} = \lambda/2$. The effective refractive index is assumed to be $n_{\mathrm{eff}} = 1.44$ \cite{10945421}, and the propagation loss of the dielectric waveguide is set to 0.1 dB/m~\cite{pozar1998microwave}. The total power radiated ratio for each waveguide is constrained by $P_{{\rm L},m} = 0.9P_{m}$.

For the multicast configuration, the carrier frequency is set to $f_c = 28$ GHz, and the noise power at each user is assumed to be $\sigma^2 = -90$ dBm. Unless otherwise specified, the users are assumed to be uniformly distributed within the service region. All numerical results are obtained by averaging over $1000$ independent random channel realizations.

For the algorithm configuration, the smoothing parameter in~\eqref{smoothing} is set to $\tau = 100$. The step size of the PGD algorithm in~\eqref{final_p} follows a diminishing rule $\alpha^{(t)} = 1/\sqrt{t+1}$. In~\eqref{update_mu}, we set the constants as $\rho_c = 1$ and $\rho_{\mu} = 0.02$ for controlling the convergence accuracy of the PAGD algorithm, while the parameters for the diminishing step size in~\eqref{update_xi} are configured as $\rho_p = 1$, $\rho_t = 10$, and $\rho_{\eta} = 0.01$. The stopping tolerance for both the outer AO framework and the inner PGD and PAGD iterations is set to $\epsilon = 10^{-4}$.

\subsection{Baseline Architectures}
We compare the proposed PASS with two representative MIMO architectures: \textit{conventional MIMO} and \textit{massive MIMO}. The specific configurations of these baselines are described as follows:

\begin{itemize}
    \item[(1)] \textbf{\textit{Conventional MIMO}}: This baseline is equipped with $N$ antennas centered at the position $[D_{\rm x}/2, 0, h]^{\rm T}$, and uniformly spaced along the $y$-axis with an equal interval $d_{\rm y}$. Each antenna is connected to a dedicated RF chain, and the system adopts fully digital signal processing.
    
    \item[(2)] \textbf{\textit{Massive MIMO}}: This baseline comprises a large-scale array of $MN$ antennas, each with its own RF chain, and also employs fully digital processing. the antenna arrays are configured as half-wavelength-spaced uniform linear arrays, centered at the location $[D_{\rm x}/2, 0, h]^{\rm T}$ and aligned along the $y$-axis within the square service region. Owing to the large number of RF chains, this configuration incurs significantly higher hardware complexity and energy consumption compared to both conventional MIMO and the proposed PASS.
\end{itemize}

For both baselines, the multicast beamforming is designed using the SOCP method mentioned in~\eqref{eq:Problem-F-socp-final}. 

\subsection{Convergence of the AO-Based Joint Beamforming}
\begin{figure}[!t]
\centering
\includegraphics[height=0.3\textwidth]{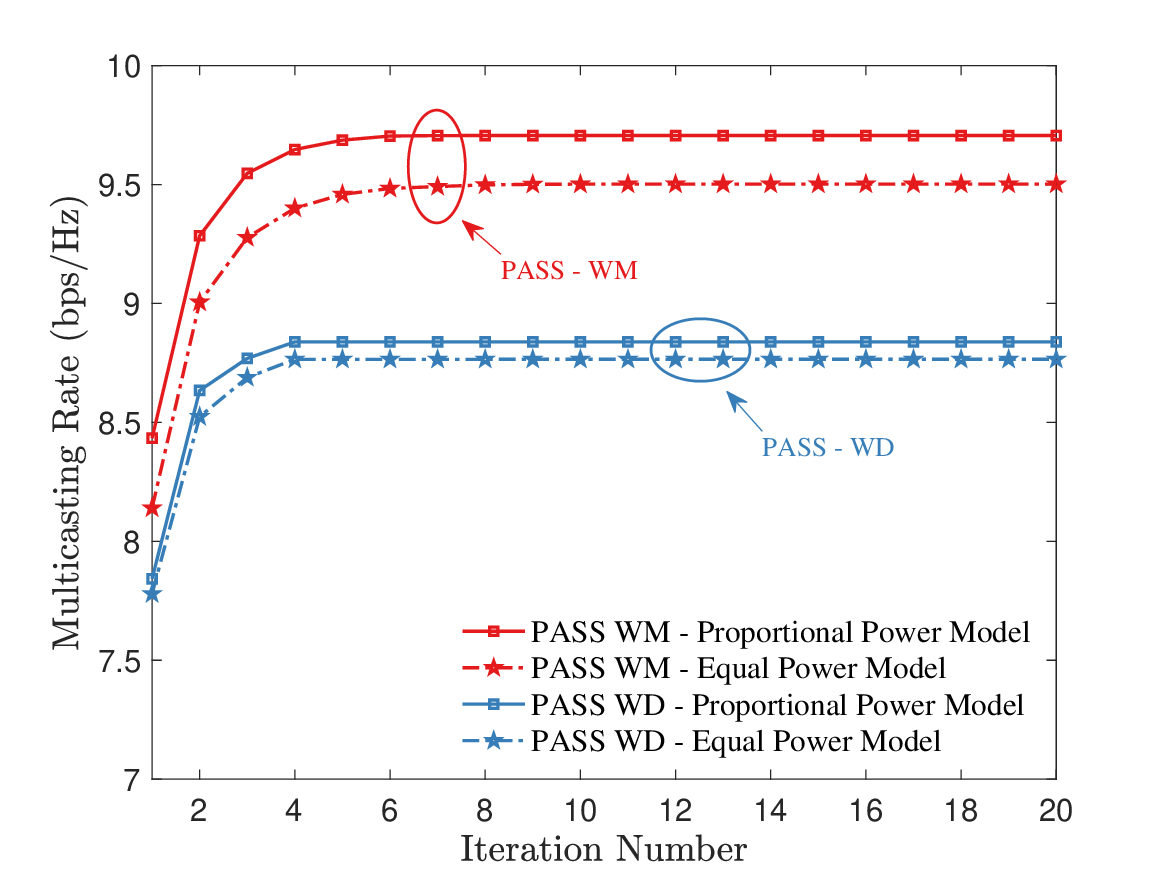}
\caption{Convergence behavior of the proposed AO algorithms.}
\label{Convergence}
\vspace{-5pt}
\end{figure}
We first evaluate the convergence behavior of the Algorithm~\ref{algo:AO} and Algorithm~\ref{algo:MM-PAGD-AO}, where the pinching beamformer and transmit beamformer are jointly optimized under the AO framework. In simulation, both the pinching and transmit beamformers are initialized randomly. As shown in {\figurename}~{\ref{Convergence}}, the multicast rate achieved by the proposed algorithm increases rapidly with the number of iterations. This result confirms the convergence of the method and its effectiveness in joint beamforming design. 

Notably, the convergence under the WD architecture is significantly faster than that under the WM architecture. This observation can be attributed to the simpler optimization structure of the WD setup: while the WM architecture requires solving a more complex transmit beamforming problem, the WD architecture only involves power allocation across independent waveguides. This further highlights the effectiveness and efficiency of the proposed LSE-PGD method. Furthermore, the rapid convergence in the WD architecture implies that power allocation has a relatively minor impact on performance improvement once the PA locations are fixed, which indicates that the overall system performance under WD architectures is more sensitive to pinching beamforming design than to precise transmit power allocation.

\subsection{Multicast Rate vs. the Transmit Power}
\begin{figure}[!t]
\centering
\includegraphics[height=0.3\textwidth]{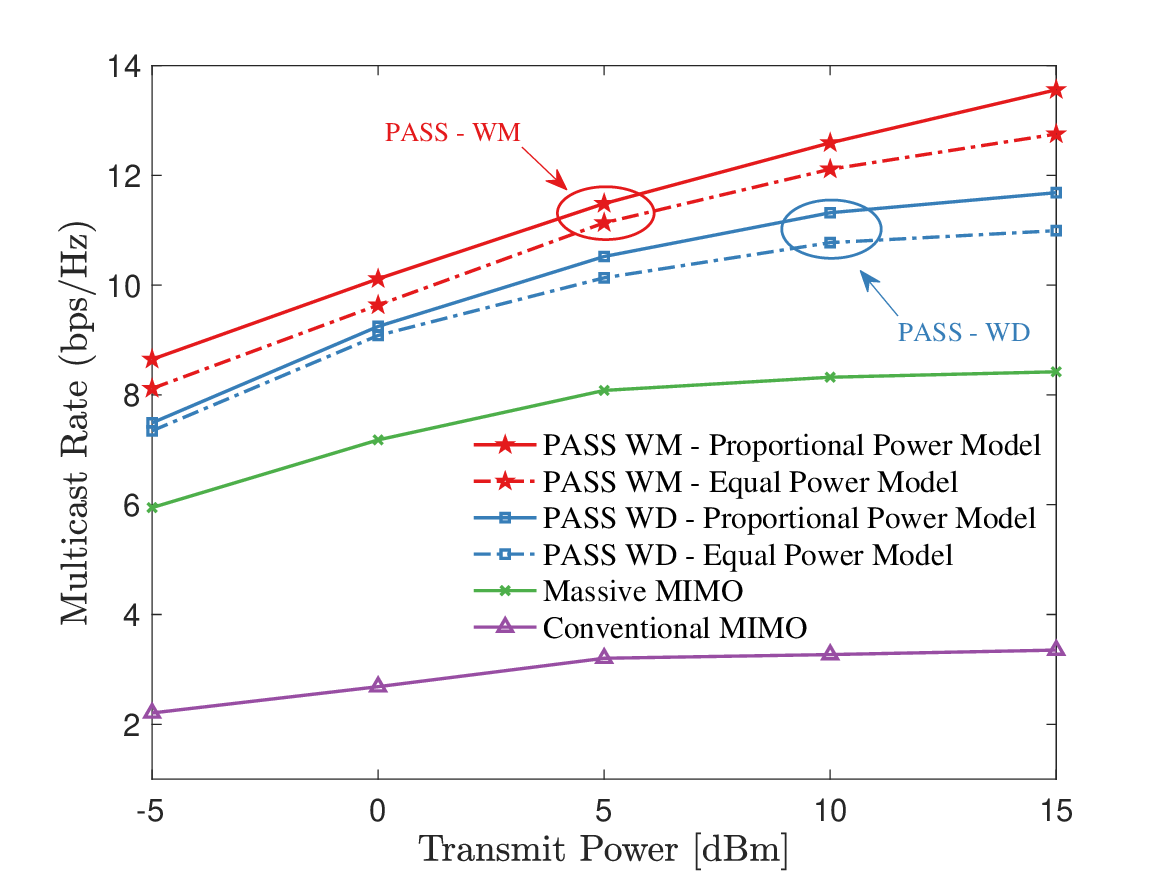}
\caption{Multicast rate versus the transmit power with $M = 4$, $N = 8$, $G = 4$, $K = 3$, $D_{\rm x} = 10$ m, and $D_{\rm y} = 6$ m.}
\label{Power}
\vspace{-5pt}
\end{figure}
{\figurename}~{\ref{Power}} shows the multicast rate as a function of transmit power. As expected, the multicast rate of all PASS-based schemes increases monotonically with transmit power. As shown in Fig.~\ref{Power}, the proportional power model consistently outperforms the equal power model under both the WD and WM architectures.  The superior performance of the proportional power model is attributed to its ability to allocate power according to the waveguide's natural propagation characteristics. Specifically, by assigning higher radiation power to PAs located closer to the waveguide's feed point, this model effectively reduces the cumulative in-waveguide propagation loss.

Furthermore, the multicast performance gain by PASS over conventional fixed-location antenna systems becomes more pronounced at higher transmit power levels, especially for the WM architecture. This observation highlights the growing significance of inter-group interference in the high SINR region, where system capacity tends to be interference-limited. In contrast, PASS leverages the spatial flexibility of large-scale waveguide deployment to suppress inter-group interference more effectively, which leads to a significant improvement in multicast performance.

\subsection{Multicast Rate vs. the Antenna Number}
\begin{figure}[!t]
\centering
\includegraphics[height=0.3\textwidth]{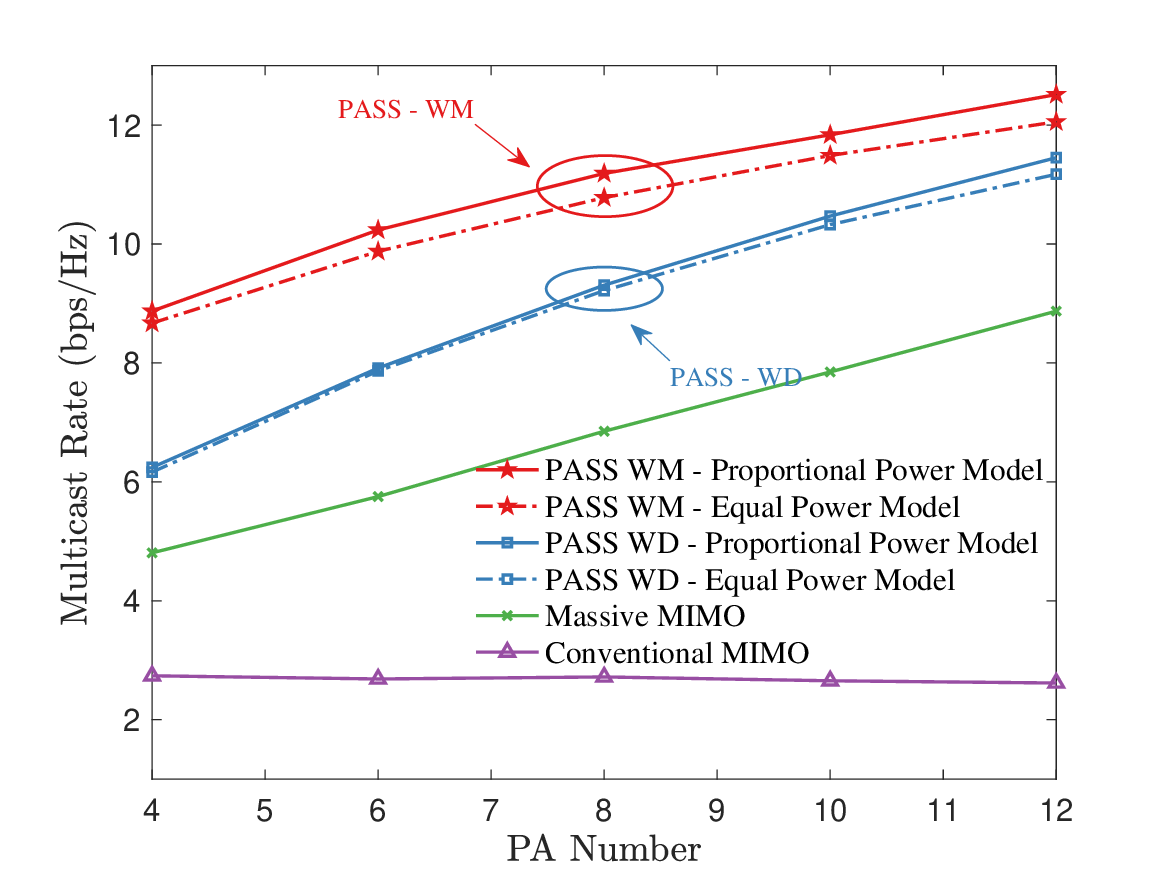}
\caption{Multicast rate versus the number of PA with $M = 4$, $G = 4$, $K = 3$, $P_{\rm t} = 0$ dBm, ${D_{\rm x} = 10}$ m and $D_{\rm y} = 6$ m.}
\label{panumber}
\vspace{-5pt}
\end{figure}
{\figurename}~{\ref{panumber}} illustrates the multicast rate as a function of the number of antennas, $N$. The results indicate that the multicast performance of PASS improves with an increasing number of PAs. This is attributed to the enhanced ability of PASS to concentrate transmit energy more effectively, which improves the multicast transmission rate.
This trend is consistent with the analytical results in \cite{ouyang2025array}, which showed that increasing the number of PAs yields a higher array gain. Moreover, as the number of PAs increases, the performance gap between the WM and WD architectures gradually narrows. This trend can be attributed to the growing dominance of the pinching beamforming gain in determining the overall system performance. With more PAs available, both architectures exhibit enhanced capability in focusing transmit energy toward intended directions, which reduces the relative multiplexing advantage previously held by the WM structure. As a result, the system becomes increasingly dependent on the effectiveness of the pinching beamforming strategy itself, rather than on the underlying transmission architecture.

In massive MIMO systems, the multicast rate can also be improved by increasing the number of antennas. However, this requires a proportional increase in RF chains and results in higher hardware complexity and energy consumption. In contrast, PASS maintains a relatively simple architecture while achieving significant performance gains. These results confirm that PASS offers an efficient solution for multicast communication.

\subsection{Multicast Rate vs. the User Number}
\begin{figure}[!t]
\centering
\includegraphics[height=0.3\textwidth]{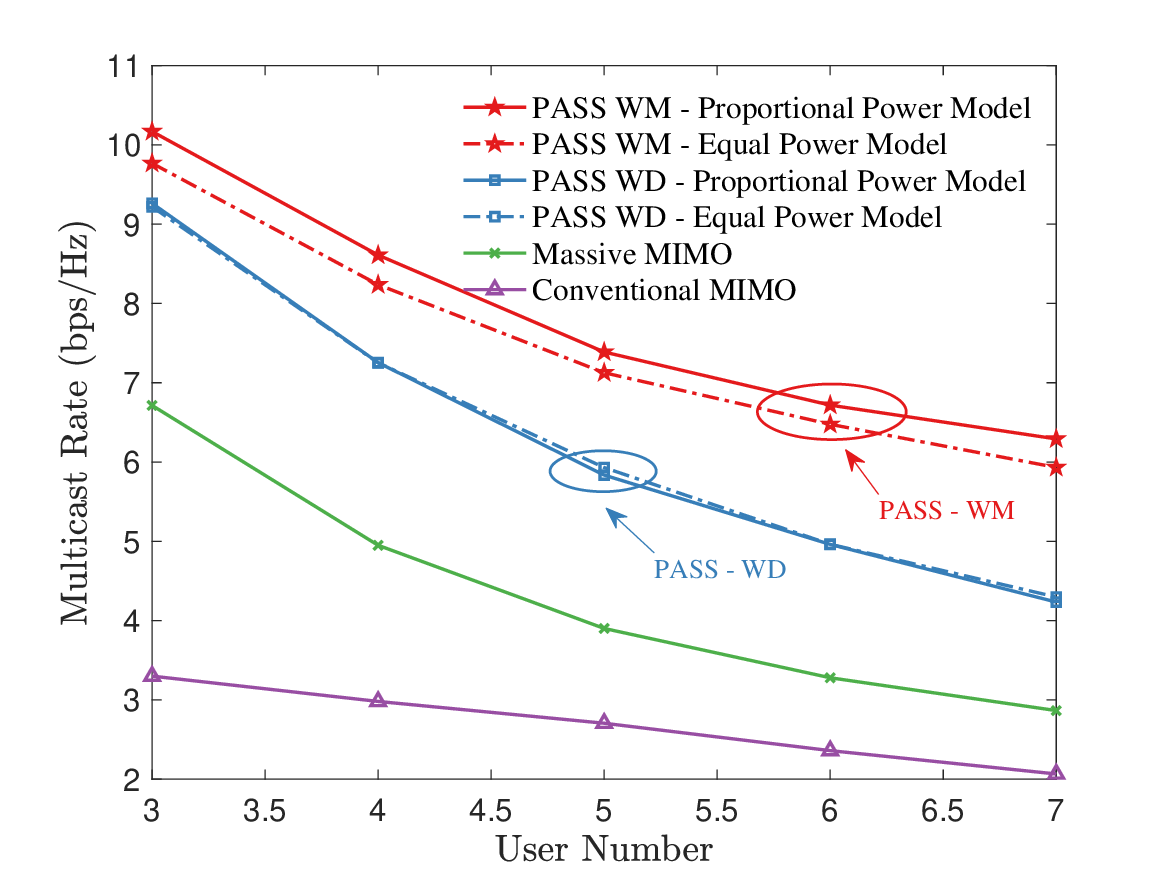}
\caption{Multicast rate versus the number of user with $M = 4$, $N = 8$, $G = 4$, $P_{\rm t} = 0$ dBm, ${D_{\rm x} = 10}$ m, and $D_{\rm y} = 6$ m.}
\label{uenumber}
\vspace{-5pt}
\end{figure}
{\figurename}~{\ref{uenumber}} illustrates the achievable multicast rate as a function of the number of users. Across all settings, PASS outperforms both conventional MIMO and massive MIMO systems. Furthermore, the performance gap between the WM and WD architectures becomes more significant as the number of users increases. This trend is primarily due to the limited interference management capability of the WD architecture, which does not employ baseband transmit beamforming. As user density grows, the system becomes increasingly susceptible to inter-group interference, which leads to substantial degradation in multicast performance. 

In contrast, the WM architecture applies baseband beamforming to exploit spatial degrees of freedom. This enables more effective interference suppression and helps maintain higher multicast rates under denser user deployments. Notably, even without baseband transmit beamforming, the WD architecture still achieves significantly better performance than traditional massive MIMO systems. This result highlights the efficiency of PASS in multicast communications with lower complexity and reduced hardware requirements. 

\subsection{Multicast Rate vs. the Group Number}
\begin{figure}[!t]
\centering
\includegraphics[height=0.3\textwidth]{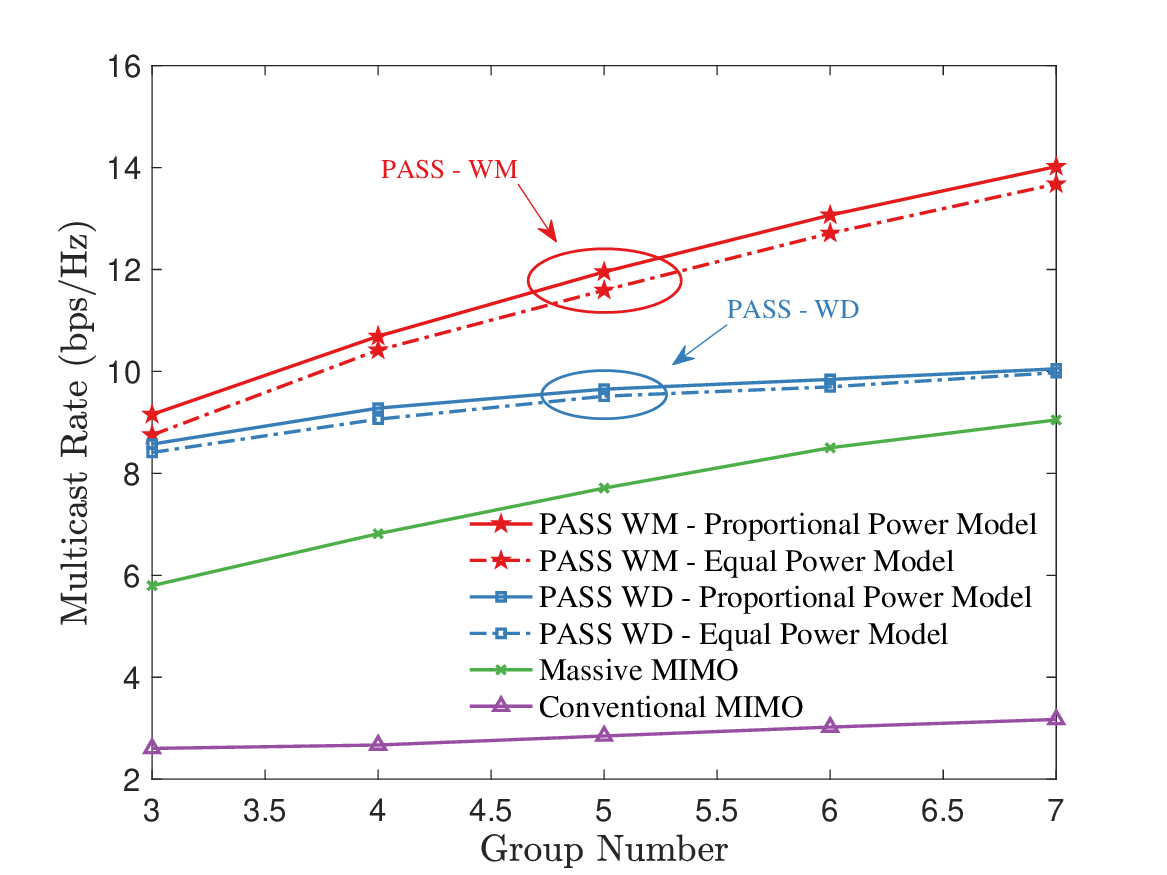}
\caption{Multicast rate versus the number of group with $M = 4$, $N = 8$, $K = 3$, $P_{\rm t} = 0$ dBm, ${D_{\rm x} = 10}$ m, and $D_{\rm y} = 6$ m.}
\label{group}
\vspace{-5pt}
\end{figure}
{\figurename}~\ref{group} illustrates the multicast rate as a function of the number of multicast groups. In this simulation, the number of groups is set equal to the number of waveguides, which ensures waveguide division multiple access for WD architecture. As the number of multicast groups increases, inter-group interference becomes more severe due to overlapping user spatial region. In this context, PASS offers substantial advantages by enabling dynamic adjustment of PA positions along the waveguides. This spatial flexibility promotes higher channel diversity and enables better alignment between transmit beamforming and group-specific channel directions, which can effectively suppress inter-group interference.

Moreover, it is noteworthy that the performance of the WM architecture scales more favorably than that of the WD counterpart as the group number increases. This is consistent with the result observed in {\figurename}~\ref{uenumber}, and can be attributed to the superior interference management capabilities provided by the baseband transmit beamforming in WM. In contrast, the WD architecture lacks coordinated digital processing, which makes it less capable of handling denser multigroup scenarios.

\subsection{Multicast Rate vs. the Side Length $D_{\rm x}$}
\begin{figure}[!t]
\centering
\includegraphics[height=0.3\textwidth]{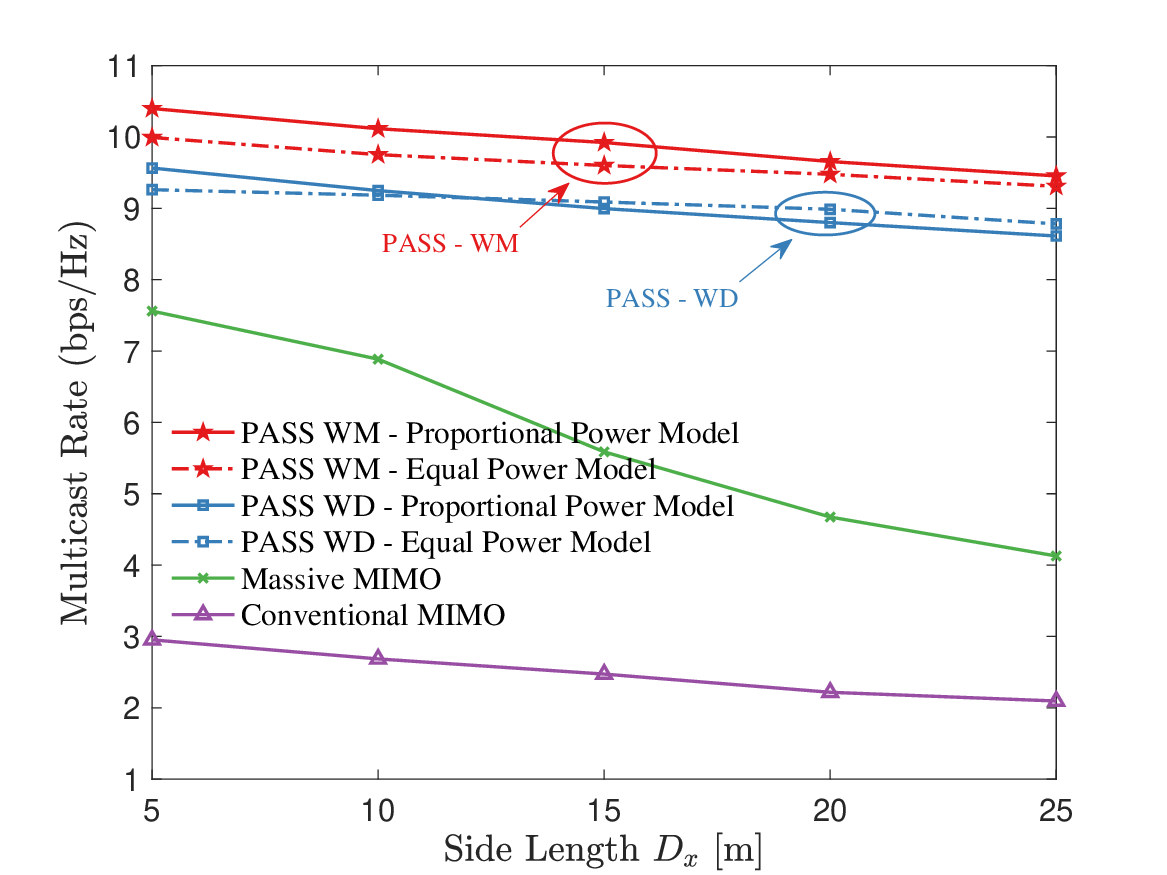}
\caption{Multicast rate versus the side length $D_{\rm x}$ with $M = 4$, $N = 8$, $G = 4$, $K = 3$, $P_{\rm t} = 0$ dBm, and $D_{\rm y} = 6$ m.}
\label{sidelength}
\vspace{-5pt}
\end{figure}

\begin{figure}[!t]
\centering
\includegraphics[height=0.3\textwidth]{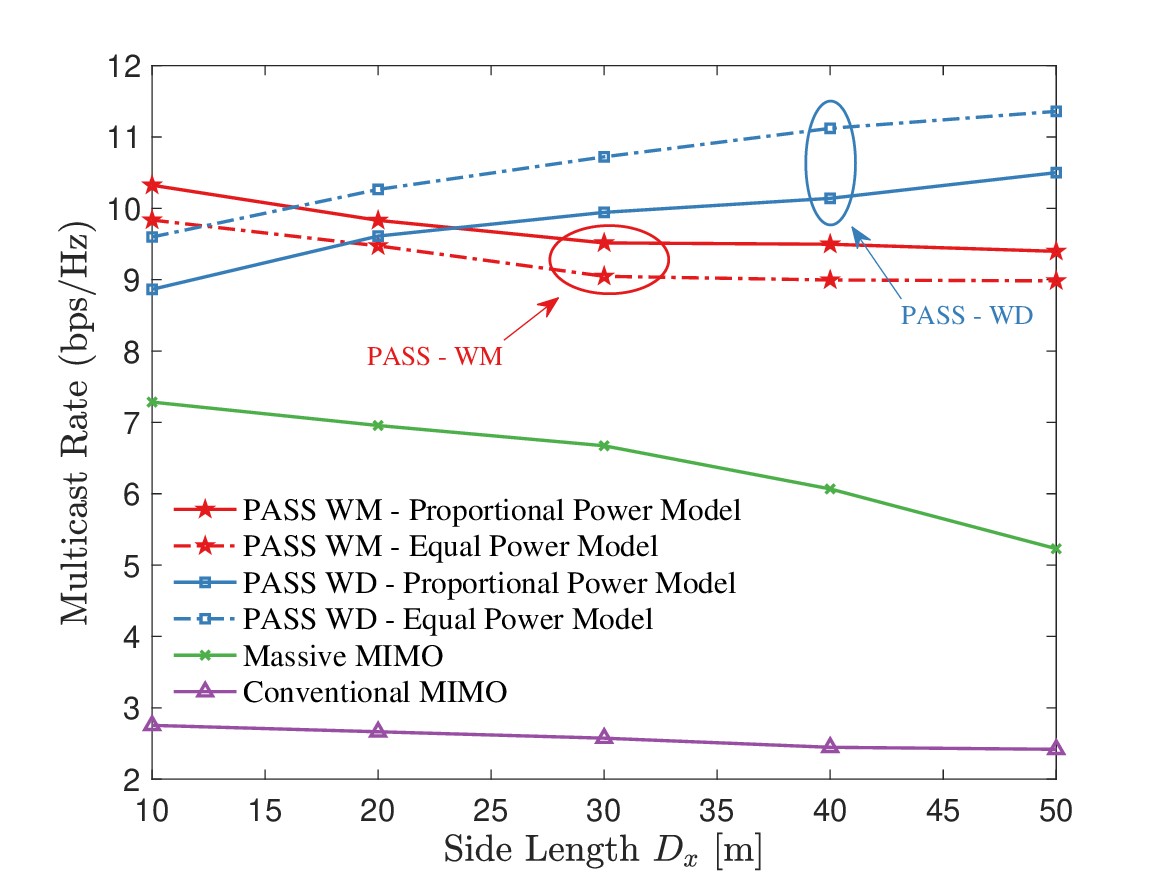}
\caption{Multicast rate versus the side length $D_{\rm x}$ under the geographically separated group distribution with $M = 4$, $N = 8$, $G = 4$, $K = 3$, $P_{\rm t} = 0$ dBm, and $D_{\rm y} = 6$ m.}
\label{sidelength_iso}
\vspace{-5pt}
\end{figure}

{\figurename} \ref{sidelength} and {\figurename} \ref{sidelength_iso} illustrate the multicast rate performance as a function of the deployment region side length $D_{\mathrm{x}}$ under two different user distribution scenarios. In {\figurename} \ref{sidelength}, users are uniformly distributed within the service region, while in {\figurename} \ref{sidelength_iso}, multicast groups are deployed in non-overlapping regions along the $x$-axis.

As observed in {\figurename}~{\ref{sidelength}}, both the WD and WM architectures of PASS consistently outperform conventional MIMO and massive MIMO baselines across all values of $D_{\mathrm{x}}$. It is noteworthy that even though massive MIMO employs spatial multiplexing to exploit available spatial degrees of freedom, its performance remains inferior to the WD architecture, despite operating without baseband multiplexing. This demonstrates the spatial adaptability offered by PASS.
In addition, while the multicast rate of all schemes decreases as the deployment region expands, the degradation for PASS is significantly slower. This robustness arises from the fact that multicast performance is fundamentally constrained by the worst-case user in each group. In conventional fixed-location antenna systems, enlarging the coverage region leads to increased link distances and higher path loss. In contrast, PASS allows PAs to dynamically follow user locations, thereby maintaining relatively stable LoS link distances even as $D_{\mathrm{x}}$ increases. These results indicate that PASS is particularly well-suited for large-scale deployments, where its spatial reconfigurability ensures robust and scalable multicast performance.

Under the geographically isolated group distribution as shown in {\figurename}~{\ref{sidelength_iso}}, the WD architecture notably outperforms the WM architecture, which differs significantly from previous uniformly overlapping scenarios. This reversal arises primarily due to the spatial separation of multicast user groups. Specifically, while the WM architecture exploits multiplexing at the baseband to share waveguide resources efficiently, the resulting inter-group interference becomes detrimental when users are distinctly partitioned into isolated spatial clusters. In contrast, the WD architecture benefits significantly from the isolated spatial distribution, as each waveguide independently serves geographically separated groups, which effectively eliminates inter-group interference and fully capitalizes on the pinching beamforming gain. This finding shows that the WD architecture is particularly well-suited for scenarios involving geographically separated user groups, as also discussed in~\cite{liu2025pass}.

\section{Conclusion}
\label{conclusion}
This article investigated multigroup multicast design for PASS. We first refined the power radiation model. Based on this, we proposed a PASS-enabled multigroup multicast framework applicable to both WD and WM transmission architectures.
For the WD architecture, we designed pinching beamforming by adopting an element-wise sequential strategy and optimized the transmit power allocation using an LSE-PGD algorithm. 
For the WM architecture, we constructed an MM surrogate function and derived the optimal transmit beamformer structure, based on which we developed an MM-PAGD algorithm for transmit beamforming and an MM-based element-wise sequential method for pinching beamforming.
Numerical results confirmed that PASS significantly enhances multicast performance over conventional fixed-location antenna systems, and that PASS exhibits robustness with increasing service region size. Furthermore, it was shown that WM provides superior interference management in dense user deployments, while WD is more effective when multicast groups are geographically separated. These results validate PASS as a flexible and low-complexity solution for next-generation physical-layer multicast communications, particularly well-suited for large-scale and dense user deployments.

\bibliographystyle{IEEEtran} 
\bibliography{reference}    

\begin{thebibliography}{10}
\providecommand{\url}[1]{#1}
\csname url@samestyle\endcsname
\providecommand{\newblock}{\relax}
\providecommand{\bibinfo}[2]{#2}
\providecommand{\BIBentrySTDinterwordspacing}{\spaceskip=0pt\relax}
\providecommand{\BIBentryALTinterwordstretchfactor}{4}
\providecommand{\BIBentryALTinterwordspacing}{\spaceskip=\fontdimen2\font plus
\BIBentryALTinterwordstretchfactor\fontdimen3\font minus \fontdimen4\font\relax}
\providecommand{\BIBforeignlanguage}[2]{{%
\expandafter\ifx\csname l@#1\endcsname\relax
\typeout{** WARNING: IEEEtran.bst: No hyphenation pattern has been}%
\typeout{** loaded for the language `#1'. Using the pattern for}%
\typeout{** the default language instead.}%
\else
\language=\csname l@#1\endcsname
\fi
#2}}
\providecommand{\BIBdecl}{\relax}
\BIBdecl

\bibitem{6542746}
A.~Adhikary, J.~Nam, J.-Y. Ahn, and G.~Caire, ``Joint spatial division and multiplexing---the large-scale array regime,'' \emph{IEEE Trans. Inf. Theory}, vol.~59, no.~10, pp. 6441--6463, 2013.

\bibitem{9424177}
Y.~Liu, X.~Liu, X.~Mu, T.~Hou, J.~Xu, M.~Di~Renzo, and N.~Al-Dhahir, ``Reconfigurable intelligent surfaces: Principles and opportunities,'' \emph{IEEE Commun. Surv. Tutor.}, vol.~23, no.~3, pp. 1546--1577, 2021.

\bibitem{10753482}
W.~K. New, K.-K. Wong, H.~Xu, C.~Wang, F.~R. Ghadi, J.~Zhang, J.~Rao, R.~Murch, P.~Ram{\'\i}rez-Espinosa, D.~Morales-Jimenez, C.-B. Chae, and K.-F. Tong, ``A tutorial on fluid antenna system for 6{G} networks: Encompassing communication theory, optimization methods and hardware designs,'' \emph{IEEE Commun. Surv. Tutor.}, early access, 2024.

\bibitem{9264694}
K.-K. Wong, A.~Shojaeifard, K.-F. Tong, and Y.~Zhang, ``Fluid antenna systems,'' \emph{IEEE Trans. Wirel. Commun.}, vol.~20, no.~3, pp. 1950--1962, 2021.

\bibitem{10286328}
L.~Zhu, W.~Ma, and R.~Zhang, ``Movable antennas for wireless communication: Opportunities and challenges,'' \emph{IEEE Commun. Mag.}, vol.~62, no.~6, pp. 114--120, 2024.

\bibitem{10906511}
L.~Zhu, W.~Ma, W.~Mei, Y.~Zeng, Q.~Wu, B.~Ning, Z.~Xiao, X.~Shao, J.~Zhang, and R.~Zhang, ``A tutorial on movable antennas for wireless networks,'' \emph{IEEE Commun. Surv. Tutor.}, early access, 2025.

\bibitem{fukuda2022pinching}
A.~Fukuda, H.~Yamamoto, H.~Okazaki, Y.~Suzuki, and K.~Kawai, ``Pinching antenna: Using a dielectric waveguide as an antenna,'' \emph{NTT DOCOMO Tech. J.}, vol.~23, no.~3, pp. 5--12, 2022.

\bibitem{liu2025pass}
Y.~Liu, Z.~Wang, X.~Mu, C.~Ouyang, and X.~Xu, ``Pinching antenna systems ({PASS}): Architecture designs, opportunities, and outlook,'' \emph{arXiv preprint, arXiv: 2501.18409}, 2025.

\bibitem{10945421}
Z.~Ding, R.~Schober, and H.~Vincent~Poor, ``Flexible-antenna systems: A pinching-antenna perspective,'' \emph{IEEE Trans. Commun.}, early access, 2025.

\bibitem{9210135}
K.-K. Wong, K.-F. Tong, Z.~Chu, and Y.~Zhang, ``A vision to smart radio environment: Surface wave communication superhighways,'' \emph{IEEE Wirel. Commun.}, vol.~28, no.~1, pp. 112--119, 2021.

\bibitem{10643519}
H.~Liu, W.~K. New, H.~Xu, Z.~Chu, K.-F. Tong, K.-K. Wong, and Y.~Zhang, ``Path loss and surface impedance models for surface wave-assisted wireless communication system,'' \emph{IEEE Access}, vol.~12, pp. 125\,786--125\,799, 2024.

\bibitem{10742352}
Z.~Chu, K.-F. Tong, K.-K. Wong, C.-B. Chae, and C.~Hou~Chan, ``On propagation characteristics of reconfigurable surface wave platform: Simulation and experimental verification,'' \emph{IEEE Access}, vol.~12, pp. 168\,744--168\,754, 2024.

\bibitem{wang2025modeling}
Z.~Wang, C.~Ouyang, X.~Mu, Y.~Liu, and Z.~Ding, ``Modeling and beamforming optimization for pinching-antenna systems,'' \emph{arXiv preprint, arXiv: 2502.05917}, 2025.

\bibitem{xu2025pass}
X.~Xu, X.~Mu, Z.-J. Wang, Y.~Liu, and A.~Nallanathan, ``Pinching-antenna systems ({PASS}): Power radiation model and optimal beamforming design,'' \emph{arXiv preprint, arXiv: 2505.00218}, 2025.

\bibitem{10912473}
K.~Wang, Z.~Ding, and R.~Schober, ``Antenna activation for {NOMA} assisted pinching-antenna systems,'' \emph{IEEE Commun. Lett.}, vol.~14, no.~5, pp. 1526--1530, 2025.

\bibitem{10976621}
D.~Tyrovolas, S.~A. Tegos, P.~D. Diamantoulakis, S.~Ioannidis, C.~K. Liaskos, and G.~K. Karagiannidis, ``Performance analysis of pinching-antenna systems,'' \emph{IEEE Trans. Cogn. Commun. Netw.}, early access, 2025.

\bibitem{ouyang2025array}
C.~Ouyang, Z.~Wang, Y.~Liu, and Z.~Ding, ``Array gain for pinching-antenna systems ({PASS}),'' \emph{IEEE Commun. Lett.}, vol.~29, no.~6, pp. 1471--1475, 2025.

\bibitem{10909665}
S.~A. Tegos, P.~D. Diamantoulakis, Z.~Ding, and G.~K. Karagiannidis, ``Minimum data rate maximization for uplink pinching-antenna systems,'' \emph{IEEE Commun. Lett.}, vol.~14, no.~5, pp. 1516--1520, 2025.

\bibitem{hou2025uplink}
T.~Hou, Y.~Liu, and A.~Nallanathan, ``On the performance of uplink pinching antenna systems ({PASS}),'' \emph{arXiv preprint, arXiv: 2502.12365}, 2025.

\bibitem{10896748}
Y.~Xu, Z.~Ding, and G.~K. Karagiannidis, ``Rate maximization for downlink pinching-antenna systems,'' \emph{IEEE Commun. Lett.}, vol.~14, no.~5, pp. 1431--1435, 2025.

\bibitem{11016750}
X.~Xie, F.~Fang, Z.~Ding, and X.~Wang, ``A low-complexity placement design of pinching-antenna systems,'' \emph{IEEE Commun. Lett.}, early access, 2025.

\bibitem{bereyhi2025downlink}
A.~Bereyhi, S.~Asaad, C.~Ouyang, Z.~Ding, and H.~V. Poor, ``Downlink beamforming with pinching-antenna assisted {MIMO} systems,'' in \emph{Proc. IEEE ICC Workshops}, 2025.

\bibitem{guo2025gpass}
J.~Guo, Y.~Liu, and A.~Nallanathan, ``{GPASS}: Deep learning for beamforming in pinching-antenna systems ({PASS}),'' \emph{arXiv preprint, arXiv: 2502.01438}, 2025.

\bibitem{xu2025joint}
X.~Xu, X.~Mu, Y.~Liu, and A.~Nallanathan, ``Joint transmit and pinching beamforming for pinching antenna systems ({PASS}): Optimization-based or learning-based?'' \emph{arXiv preprint, arXiv: 2502.08637}, 2025.

\bibitem{zhao2025wdma}
J.~Zhao, X.~Mu, K.~Cai, Y.~Zhu, and Y.~Liu, ``Waveguide division multiple access for pinching-antenna systems ({PASS}),'' \emph{arXiv preprint, arXiv: 2502.17781}, 2025.

\bibitem{yeh2008essence}
C.~Yeh and F.~I. Shimabukuro, \emph{The Essence of Dielectric Waveguides}.\hskip 1em plus 0.5em minus 0.4em\relax New York, NY, USA: Springer, 2008.

\bibitem{xu2001smoothing}
S.~Xu, ``Smoothing method for minimax problems,'' \emph{Computational Optimization and Applications}, vol.~20, no.~3, pp. 267--279, 2001.

\bibitem{9076830}
G.~Zhou, C.~Pan, H.~Ren, K.~Wang, and A.~Nallanathan, ``Intelligent reflecting surface aided multigroup multicast {MISO} communication systems,'' \emph{IEEE Trans. Signal Process.}, vol.~68, pp. 3236--3251, 2020.

\bibitem{mosek2015}
{MOSEK ApS}, \emph{The MOSEK Optimization Toolbox for MATLAB Manual}, \url{http://mosek.com}, Online; accessed Mar. 20, 2015, 2015, version 7.1 (revision 28).

\bibitem{fang2023optimal}
T.~Fang and Y.~Mao, ``Optimal beamforming structure and efficient optimization algorithms for generalized multi-group multicast beamforming optimization,'' \emph{arXiv preprint, arXiv: 2312.16559}, 2023.

\bibitem{NedicBertsekas2001}
A.~Nedi{\'c} and D.~P. Bertsekas, ``Incremental subgradient methods for nondifferentiable optimization,'' \emph{SIAM J. Optim.}, vol.~12, no.~1, pp. 109--138, 2001.

\bibitem{Michelot1986}
C.~Michelot, ``A finite algorithm for finding the projection of a point onto the canonical simplex of $\mathbb{R}^n$,'' \emph{J. Optim. Theory Appl.}, vol.~50, no.~1, pp. 195--200, 1986.

\bibitem{NedicOzdaglar2009}
A.~Nedi{\'c} and A.~E. Ozdaglar, ``Distributed subgradient methods for multi-agent optimization,'' \emph{IEEE Trans. Autom. Control}, vol.~54, no.~1, pp. 48--61, 2009.

\bibitem{pozar1998microwave}
D.~M. Pozar, \emph{Microwave Engineering}, 4th~ed.\hskip 1em plus 0.5em minus 0.4em\relax New York, US: Wiley, 1998.

\end{thebibliography}

\end{document}